%% file: sample-acmsmall.tex
\documentclass[acmsmall]{acmart}

\AtBeginDocument{%
  \providecommand\BibTeX{{%
    \normalfont B\kern-0.5em{\scshape i\kern-0.25em b}\kern-0.8em\TeX}}}

\setcopyright{acmcopyright}
\copyrightyear{2026}
\acmYear{2026}
\acmDOI{XXXXXXX.XXXXXXX}

\acmJournal{TOSEM}
\acmVolume{0}
\acmNumber{0}
\acmArticle{0}
\acmMonth{0}

\usepackage{longtable}
\usepackage{booktabs}
\usepackage{tabularx}
\usepackage{array}
\usepackage{multirow}
\usepackage{makecell}
\usepackage{dcolumn}
\usepackage{colortbl}
\usepackage{adjustbox}
\usepackage{tabu}
\usepackage{supertabular}
\usepackage{diagbox}
\usepackage{siunitx}
\newcolumntype{Y}{>{\raggedright\arraybackslash}X}

\usepackage{graphicx}
\usepackage{subcaption}
\usepackage{wrapfig}
\usepackage{placeins}

\usepackage[table]{xcolor}
\definecolor{codebg}{RGB}{243,226,212}
\definecolor{codekw}{RGB}{23,49,62}
\definecolor{codecomment}{RGB}{197,176,205}
\definecolor{codeem}{RGB}{65,94,114}
\definecolor{softblue}{RGB}{222,235,247}
\definecolor{softred}{RGB}{248,215,218}
\definecolor{softgold}{RGB}{247,237,203}
\definecolor{codegreen}{rgb}{0.25,0.49,0.28}
\definecolor{codepurple}{rgb}{0.58,0.0,0.52}
\definecolor{codered}{rgb}{0.68,0.13,0.12}
\definecolor{bglight}{HTML}{F7F7F7}
\definecolor{kwblue}{HTML}{0060C0}
\definecolor{strgreen}{HTML}{267F00}
\definecolor{commentgray}{HTML}{808080}
\definecolor{enumorange}{HTML}{B85C00}
\definecolor{codegray}{rgb}{0.5,0.5,0.5}

\usepackage{mathtools}
\usepackage{bbm}

\newcommand{\Ind}[1]{\mathbf{1}_{\mathrm{#1}}}
\newcommand{\tplus}{\mathbin{\text{+}}}

\usepackage{microtype}
\usepackage[T1]{fontenc}
\usepackage[utf8]{inputenc}
\usepackage{ragged2e}
\usepackage{enumitem}
\usepackage[normalem]{ulem}
\usepackage{pifont}
\usepackage{comment}

\usepackage{mdframed}
\usepackage[most]{tcolorbox}
\newtcolorbox{findingbox}[1][]{ #1}
\newtcolorbox{box2}[1][]{
  title=#1,
  enhanced,
  colback=gray!20,
  frame hidden,
  colbacktitle=gray,
  boxed title style={colframe=gray},
  top=0.05cm,
  bottom=0.05cm,
  left=0.15cm,
  right=0.15cm,
  enlarge top by=0.5cm,
  enlarge bottom by=0.3cm,
  attach boxed title to top left={xshift=3mm,yshift=-3mm,yshifttext=-1mm},
  borderline west={4pt}{0pt}{blue!50!black},
  breakable
}

\usepackage{algpseudocode}
\RequirePackage[linesnumbered,ruled]{algorithm2e}
\SetKw{Break}{break}
\SetKwRepeat{Do}{do}{until}
\SetKwInOut{Input}{Input}
\SetKwInOut{Output}{Output}

\usepackage{listings}
\usepackage{fancyvrb}
\lstdefinelanguage{json}{
    basicstyle=\ttfamily\small,
    showstringspaces=false,
    breaklines=true,
    frame=single,
    morestring=[b]",
    stringstyle=\color{black},
    literate=
     *{0}{{{\color{black}0}}}{1}
      {1}{{{\color{black}1}}}{1}
      {2}{{{\color{black}2}}}{1}
      {3}{{{\color{black}3}}}{1}
      {4}{{{\color{black}4}}}{1}
      {5}{{{\color{black}5}}}{1}
      {6}{{{\color{black}6}}}{1}
      {7}{{{\color{black}7}}}{1}
      {8}{{{\color{black}8}}}{1}
      {9}{{{\color{black}9}}}{1}
      {:}{{{\color{black}{:}}}}{1}
      {,}{{{\color{black}{,}}}}{1}
      {\{}{{{\color{black}{\{}}}}{1}
      {\}}{{{\color{black}{\}}}}}{1}
      {[}{{{\color{black}{[}}}}{1}
      {]}{{{\color{black}{]}}}}{1},
}
\lstdefinelanguage{JavaScript}{
  keywords={const, let, var, function, return, if, else, for, while, break, continue},
  keywordstyle=\color{blue}\bfseries,
  ndkeywords={class, export, boolean, throw, implements, import, this},
  ndkeywordstyle=\color{blue}\bfseries,
  identifierstyle=\color{black},
  sensitive=true,
  comment=[l]{//},
  morecomment=[s]{/*}{*/},
  commentstyle=\color{codegreen}\itshape,
  stringstyle=\color{codepurple},
  morestring=[b]',
  morestring=[b]"
}
\lstdefinestyle{jsoncode}{
  language=json,
  basicstyle=\ttfamily\small,
  keywordstyle=\color{blue}\bfseries,
  stringstyle=\color{red},
  commentstyle=\color{gray}\itshape,
  numbers=left,
  numberstyle=\tiny\color{gray},
  stepnumber=1,
  numbersep=5pt,
  frame=single,
  breaklines=true,
  breakatwhitespace=true,
  tabsize=2,
  showstringspaces=false
}
\lstdefinestyle{code}{
  backgroundcolor=\color{white},
  basicstyle=\ttfamily\footnotesize,
  breaklines=true,
  frame=single,
  framerule=0.4pt,
  rulecolor=\color{black!50},
  captionpos=b,
  numbers=none
}
\lstdefinestyle{yamlstyle}{
  language={},
  basicstyle=\ttfamily\scriptsize,
  backgroundcolor=\color{bglight},
  frame=single,
  rulecolor=\color{gray!40},
  xleftmargin=3pt, xrightmargin=3pt,
  aboveskip=4pt, belowskip=2pt,
  moredelim=[s][\color{kwblue}\bfseries]{-\ name}{:},
  moredelim=[s][\color{kwblue}\bfseries]{\ \ type}{:},
  moredelim=[s][\color{kwblue}\bfseries]{required}{:},
  moredelim=[s][\color{kwblue}\bfseries]{enum}{:},
  moredelim=[s][\color{kwblue}\bfseries]{schema}{:},
  moredelim=[s][\color{kwblue}\bfseries]{in}{:},
  moredelim=[s][\color{strgreen}]{"}{"},
  moredelim=[l][\color{commentgray}]{\#},
  showstringspaces=false,
  columns=flexible,
}
\lstdefinestyle{gostyle}{
  language={},
  basicstyle=\ttfamily\scriptsize,
  backgroundcolor=\color{bglight},
  frame=single,
  rulecolor=\color{gray!40},
  xleftmargin=3pt, xrightmargin=3pt,
  aboveskip=4pt, belowskip=2pt,
  morekeywords={Annotations,Title,Tool,Name,Description,InputSchema,Type,Properties,
    Required,Enum,ReadOnlyHint,Schema},
  keywordstyle=\color{kwblue}\bfseries,
  moredelim=[s][\color{strgreen}]{"}{"},
  moredelim=[l][\color{commentgray}]{//},
  showstringspaces=false,
  columns=flexible,
}
\lstdefinestyle{yamlstyle2}{
  language={},
  basicstyle=\ttfamily\scriptsize,
  backgroundcolor=\color{bglight},
  frame=single,
  rulecolor=\color{gray!40},
  xleftmargin=3pt, xrightmargin=3pt,
  aboveskip=4pt, belowskip=2pt,
  showstringspaces=false,
  columns=fullflexible,
  keepspaces=true,
  breaklines=true,
  moredelim=[l][\color{commentgray}]{\#},
  moredelim=[s][\color{strgreen}]{"}{"},
  morekeywords={
    get,post,put,patch,delete,
    summary,description,operationId,
    parameters,responses,content,
    schema,type,properties,items,
    required,enum,in,name
  },
  keywordstyle=\color{kwblue}\bfseries
}
\lstdefinestyle{gostyle2}{
  language={},
  basicstyle=\ttfamily\scriptsize,
  backgroundcolor=\color{bglight},
  frame=single,
  rulecolor=\color{gray!40},
  xleftmargin=3pt, xrightmargin=3pt,
  aboveskip=4pt, belowskip=2pt,
  showstringspaces=false,
  columns=fullflexible,
  keepspaces=true,
  breaklines=true,
  moredelim=[l][\color{commentgray}]{//},
  moredelim=[s][\color{strgreen}]{"}{"},
  morekeywords={
    Annotations,Title,readOnlyHint,
    description,inputSchema,
    type,required,properties,
    query,perPage,page,
    minimum,maximum,name
  },
  keywordstyle=\color{kwblue}\bfseries
}

\usepackage{etoolbox}
\newtoggle{showloc}
\togglefalse{showloc}

\usepackage{hyperref}

\usepackage{assets/templates/iselab}

\makeatletter
\providecommand{\sf@counterlist}{}
\makeatother

\newcommand{\meriem}[1]{\textcolor{purple}{#1}}

\begin{document}

\title{From REST to MCP: An Empirical Study of API Wrapping and Automated Server Generation for LLM Agents}

\author{Meriem Mastouri}
\affiliation{%
  \institution{Grand Valley State University}
  \state{MI}
  \country{USA}
}
\email{mastourm@mail.gvsu.edu}

\author{Emna Ksontini}
\affiliation{%
  \institution{University of North Carolina Wilmington}
  \state{NC}
  \country{USA}
}
\email{ksontinie@uncw.edu}

\author{Amine Barrak}
\affiliation{%
  \institution{Oakland University}
  \state{MI}
  \country{USA}
}
\email{aminebarrak@oakland.edu}

\author{Wael Kessentini}
\affiliation{%
  \institution{DePaul University}
  \state{IL}
  \country{USA}
}
\email{wkessent@depaul.edu}

\renewcommand{\shortauthors}{Mastouri et al.}

\begin{abstract}
The Model Context Protocol (MCP) is emerging as a standard interface through which LLM agents invoke external tools, and a growing ecosystem of MCP servers now mediates access to vendor services. Most of these servers target vendors that already expose REST APIs, yet the relationship between MCP tool interfaces and the underlying API surface has not been empirically characterised. This paper presents the first large-scale study of MCP server construction. We analyse 116 official servers to determine REST reliance and integration strategies (RQ1); examine servers paired with OpenAPI specifications to quantify operation exposure, omission, and mapping patterns (RQ2); evaluate automated generation from 80 real-world OpenAPI contracts (RQ3); and assess specification repair and tool-set transformations to improve correctness and reduce complexity (RQ4). We find that 88.6\% of servers are fully or partially REST-backed, with 92\% implementing tools as bare API wrappers. MCP servers expose a median of 19\% of available operations, following systematic patterns predictable from the specification. Baseline generation succeeds for 76\% of sampled tools; automated repair raises this to 94.2\%, while filtering and regrouping reduce the median tool count per API by one-third. We release AutoMCP, an end-to-end pipeline integrating specification repair and empirically grounded tool-set transformations.
\end{abstract}

\keywords{Model Context Protocol, MCP, OpenAPI, REST API, tool-augmented LLMs, Agents, specification repair}

\maketitle

\input{Sections/Introduction}
\input{Sections/Background}

\input{Sections/RelatedWork}
\input{Sections/ResearchMethod}
\input{Sections/ResultsAndAnalysis}

\input{Sections/Limitations}

\input{Sections/Conclusion}

\bibliographystyle{ACM-Reference-Format}
\bibliography{references}

\end{document}

%% file: Sections/Introduction.tex
\section{Introduction}
\label{sec:Introduction}

Large language models are increasingly deployed as autonomous agents
that plan, reason, and act by invoking external
tools~\cite{yao2022react, schick2023toolformer}. The breadth of
services an agent can operate on is determined by the tool interfaces
available to it: machine-readable contracts that specify what each
tool does, what arguments it accepts, and how it authenticates. As
the number and diversity of target services grow, constructing and
maintaining these interfaces has become a recognised engineering
bottleneck~\cite{ chen2025llm_dev_challenges},
motivating protocol-level solutions that standardise how tools are
described, discovered, and invoked.

The Model Context Protocol (MCP)~\cite{mcp2025} is the most widely
adopted of these solutions. MCP defines a schema-driven JSON-RPC
interface through which servers expose callable tools and clients
invoke them in a uniform message format, decoupling agent logic from
service-specific implementation. Major platforms, such as GitHub, Notion,
Slack, among others, now publish official MCP servers, and public
registries~\cite{docker_mcp_catalog, mcp_so_registry} catalogue
hundreds more. 

\begin{figure*}[t]
\centering
\begin{minipage}[t]{0.47\textwidth}
\begin{lstlisting}[style=yamlstyle,
  title={\footnotesize\textbf{(a) GitHub OpenAPI---\texttt{GET /search/repositories}~\cite{github_rest_api_description}}}]
# GET /search/repositories
parameters:
  - name: q
    in: query
    required: true
    schema: { type: string }
  - name: sort
    in: query
    required: false
    schema:
      type: string
      enum: [stars, forks,
        help-wanted-issues, updated]
  - name: order
    in: query
    schema:
      type: string
      enum: [desc, asc]
\end{lstlisting}
\end{minipage}
\hfill
\begin{minipage}[t]{0.47\textwidth}
\begin{lstlisting}[style=gostyle,
  title={\footnotesize\textbf{(b) GitHub MCP tool---\texttt{search\_repositories tool}~\cite{github_mcp}}}]
Tool{
  Name: "search_repositories",
  Description: "Find GitHub repositories by name, description, or topics.",
  ReadOnlyHint: true,
  InputSchema: Schema{
    Type: "object",
    Properties: {
      "query":
        {Type: "string",
         Required: true},
      "sort":
        {Type: "string",
         Enum: ["stars","forks", "help-wanted-issues", "updated"]},
      "order":
        {Type: "string",
         Enum: ["asc","desc"]},
    },
  },
}
// Handler:
result, _, _ :=
  client.Search.Repositories(
    ctx, query, opts)
\end{lstlisting}
\end{minipage}
\caption{The \texttt{GET /search/repositories} endpoint from
GitHub's OpenAPI specification~(a) and the corresponding MCP tool
from the official GitHub MCP server~(b)~\cite{github_mcp}. Parameter
names, types, and enumerations are preserved across both
representations; the tool handler delegates directly to the vendor
REST API.}
\label{fig:motivating}
\end{figure*}

A closer look at publicly available MCP servers reveals a striking
resemblance to existing service APIs.
Figure~\ref{fig:motivating} juxtaposes a snippet from GitHub's
OpenAPI specification for the \texttt{GET /search/repositories}
endpoint with the corresponding tool definition from the official
GitHub MCP server~\cite{github_mcp}. The structural correspondence
is immediate: parameter names, types, and enumerations are preserved
almost verbatim, and the tool handler delegates directly to the
vendor REST API. More broadly, the capabilities that MCP servers
advertise, managing repositories, sending messages, querying
dashboards, closely mirror what these vendors already expose through
their public REST APIs. At the same time, the scale of the two
interfaces differs markedly. GitHub's REST API defines over
600~operations\footnote{\url{https://github.blog/news-insights/product-news/introducing-githubs-openapi-description/}},
while its MCP server exposes
51~tools\footnote{\url{https://github.com/github/github-mcp-server}}.
Slack's API comprises over
200~methods\footnote{\url{https://api.slack.com/methods}}, yet its
MCP server surfaces only
8~tools\footnote{\url{https://github.com/modelcontextprotocol/servers/tree/main/src/slack}}.
These observations raise questions that have not been empirically
investigated: what is the relationship between MCP servers and the
REST APIs offered by the same vendors? What determines which
operations are surfaced as tools and which are not? And do the
resulting tool sets reflect systematic design principles or ad~hoc
developer choices?

The answers to these questions have practical implications beyond
documentation of developer practice. Real-world REST APIs routinely
define large operation surfaces, with public corpora reporting
medians in the tens and long tails reaching hundreds of
endpoints~\cite{neumann2018analysis, golmohammadi2023testing}.
Recent evaluations consistently show that LLM tool-selection
accuracy degrades as the number of available tools grows, with
reported losses of up to
85\%~\cite{kate2025longfunceval, shi2025retrieval}. How MCP server
developers navigate the tension between API coverage and agent
usability is therefore a design problem with direct implications for
agent reliability. 

Moreover, the structural resemblance between OpenAPI endpoint definitions and 
MCP tool schemas echoes a pattern 
that software engineering has exploited before: tools such as 
Swagger Codegen and OpenAPI Generator already compile API 
specifications into clients, server stubs, and  SDKs~\cite{swagger_codegen,openapi_generator}. Whether the correspondence extends to 
MCP tool generation, and whether it survives the diversity of 
real-world APIs and the known quality issues of their 
specifications, is a question that requires  empirical investigation.

A growing body of work has begun to examine the MCP ecosystem, but
from complementary angles that leave these questions open. Security
analyses have catalogued threats across the server
lifecycle~\cite{hou2025model}. Fault
taxonomies have classified recurring defect patterns in server
implementations~\cite{taraghi2026real}. Measurement studies have
profiled ecosystem health and transport
adoption~\cite{guo2025measurement}. Description-quality studies
have identified smells in tool
metadata~\cite{hasan2026model}. However, the relationship between MCP servers and the service APIs
they target, the exposure and mapping decisions their developers
make, and whether the construction process can be systematised
remain unstudied.

This paper provides the first empirical investigation of MCP server
construction. We analyse 116~real-world MCP servers drawn from
official listings and community registries, characterising their
architecture, service-integration strategies, and configuration
practices~(\textbf{RQ1}). For the servers whose vendor APIs have
publicly available OpenAPI specifications, we examine how REST
operations are translated into MCP tools, quantifying coverage,
documenting omission patterns, and identifying the mapping
structures developers employ~(\textbf{RQ2}). Building on these
empirical findings, we investigate whether MCP server construction
can be derived from OpenAPI specifications by generating servers
from 77~real-world specifications and cataloguing the defects that
prevent faithful generation~(\textbf{RQ3}). Guided by the design
patterns uncovered in RQ1--RQ2 and the defect taxonomy from RQ3, we
introduce \textsc{AutoMCP}, an end-to-end pipeline combining
specification-driven generation, specification repair, and
empirically grounded tool-set transformations~(\textbf{RQ4}).

The remainder of this paper is organised as follows.
Section~\ref{sec:related-work} reviews related work.
Section~\ref{sec:study_design} describes the study design, dataset
construction, and the \textsc{AutoMCP} framework.
Section~\ref{sec:evaluation} presents findings across the four
research questions. Section~\ref{sec:threats-to-validity} discusses threats to
validity. Section~\ref{sec:conclusion} concludes.

\textbf{Replication Package.} All materials are available at~\cite{ESEM25_replication}.

%% file: Sections/Background.tex



%% file: Sections/RelatedWork.tex
\section{Related Work}
\label{sec:related-work}

This section reviews three areas of prior work relevant to our 
study.

\subsection{MCP Protocol-Based Tool Invocation for Language Model Agents}
\label{sec:rw-mcp}

Recent work has begun to examine the MCP ecosystem from several 
complementary perspectives. Hou~et~al.~\cite{hou2025model} analyse security threats 
across the MCP server lifecycle, showing that protocol 
standardisation alone does not produce a uniform security posture; 
risks arise from server-level implementation, credential management, 
and capability-exposure decisions. 
Hasan~et~al.~\cite{hasan2025model} conduct the first 
large-scale assessment of MCP server health, evaluating 1,899 
open-source servers for security vulnerabilities and maintainability 
concerns. Taraghi~et~al.~\cite{taraghi2026real} present the first 
fault taxonomy for MCP software, identifying five high-level 
categories of defects through analysis of server implementations 
and practitioner surveys. 
Guo~et~al.~\cite{guo2025measurement} profile the ecosystem at 
scale, collecting 17,630 entries across six registries and 
analysing 8,401 valid projects for structural health, transport 
adoption, and client--server distribution patterns. 
Hasan~et~al.~\cite{hasan2026model} study tool description 
quality, finding that 97.1\% of the 856 tools analysed contain at 
least one description smell, with 56\% failing to state their 
purpose clearly.

These studies collectively address the security, reliability, 
ecosystem health, and description quality of MCP servers. None, 
however, examines how MCP servers relate to the service APIs they 
target: what operations they expose or omit, what mapping patterns 
their developers apply, or whether their construction follows 
regularities amenable to automation. Our work addresses this gap 
through the first empirical characterisation of MCP server 
construction and its relationship to vendor REST APIs.

\subsection{LLM Tool Use and Selection at Scale}
\label{sec:rw-tool-use}

Enabling language models to invoke external tools has been a 
central focus of recent research. 
Schick~et~al.~\cite{schick2023toolformer} demonstrate that language 
models can learn to use tools autonomously through self-supervised 
training, while Yao~et~al.~\cite{yao2022react} show that 
interleaving reasoning traces with tool-invocation actions improves 
task completion in multi-step settings. These foundational results 
have motivated a line of work on scaling tool use to large and 
heterogeneous inventories. 
Patil~et~al.~\cite{patil2023gorilla} train a model to select and 
parameterise calls from a corpus of over 1,600 API endpoints, 
focusing on reducing hallucinated or malformed invocations. 
Qin~et~al.~\cite{qin2023toolllm} extend this direction to over 
16,000 real-world APIs, introducing a decision-tree search strategy 
for multi-tool planning. 
Du~et~al.~\cite{du2024anytool} propose a hierarchical agent 
architecture that decomposes complex tasks into API call sequences 
using self-reflection.

Benchmarking efforts have formalised the evaluation of tool-use 
capabilities. Li~et~al.~\cite{li2023apibank_emnlp} introduce 
API-Bank, a benchmark requiring models to select, invoke, and 
compose API calls to complete multi-step tasks. 
Chen~et~al.~\cite{chen2024t} propose T-Eval, which 
decomposes tool utilisation into fine-grained sub-skills: plan 
generation, argument completion and response interpretation, and 
evaluates each independently.

A recurring finding across this literature is that tool-selection 
accuracy degrades as the number of available tools grows. 
Kate~et~al.~\cite{kate2025longfunceval} report performance losses of 
7--85\% as tool catalogue size increases. 
Shi~et~al.~\cite{shi2025retrieval} show that even strong retrieval 
models fail on tool retrieval tasks, degrading downstream task 
completion. These results establish that the size and composition 
of the tool set presented to a model directly affect agent 
reliability.

This body of work treats the tool interface as a given input and 
studies how models interact through it. Our work is complementary: 
we study how that interface is designed and constructed, and we 
address the tool-set scalability problem from the supply side by 
introducing empirically grounded filtering and grouping 
transformations that reduce the number of tools presented to the 
model.

\subsection{REST-to-Tool Construction and Specification Quality}
\label{sec:rw-rest}

Research on connecting language models to REST APIs has focused on 
consumption: translating intent into request 
sequences~\cite{song2023restgpt}, modelling inter-endpoint 
dependencies~\cite{le2024kat, kim2024multiagent_rest_api_testing}, 
and selecting endpoints from large 
inventories~\cite{patil2023gorilla}. These approaches assume the 
tool interface exists.

On the construction side, Ni~et~al.~\cite{li2024toolfactory} 
propose ToolFactory, which synthesises callable wrappers from 
natural-language API documentation, targets non-MCP interfaces 
and does not address tool-set scalability. 
Within the MCP ecosystem, FastMCP~\cite{fastmcp}, the most starred 
adopted community framework, supports OpenAPI-to-MCP translation, 
but as a development library rather than an end-to-end pipeline. 
Developers must manually specify server base url, authentication schemes, 
and transport settings, information already encoded in the 
specification's \texttt{servers} and \texttt{securitySchemes} 
fields, making the process impractical across multiple APIs. 
FastMCP performs one-to-one endpoint-to-tool translation without 
filtering or grouping, and does not handle specification defects. 
Users report generation failures for POST endpoints with complex 
payloads~\cite{fastmcp_issue_3113, fastmcp_issue_3093}, crashes on 
circular schema references~\cite{fastmcp_issue_3242}, and 
context-window overflow from verbose tool 
descriptions~\cite{fastmcp_issue_1449}. Table~\ref{tab:rw_comparison} summarises 
the differences.
\begin{table}[t]
\centering
\small
\caption{Comparison with existing approaches.}
\label{tab:rw_comparison}
\setlength{\tabcolsep}{4pt}
\begin{tabular}{
p{4.3cm} 
>{\centering\arraybackslash}p{2.4cm} 
>{\centering\arraybackslash}p{2.4cm} 
>{\centering\arraybackslash}p{2.9cm}
}
\toprule
 & \textbf{ToolFactory}~\cite{li2024toolfactory} 
 & \textbf{FastMCP}~\cite{fastmcp} 
 & \textbf{AutoMCP} \\
\midrule
Approach type 
  & LLM-based 
  & Dev.\ library 
  & End-to-end pipeline \\
Produces MCP server 
  & $\times$ 
  & \checkmark 
  & \checkmark \\
Uses OpenAPI as input  
  & $\times$ 
  & \checkmark 
  & \checkmark \\
Auth.\ \& server config. 
  & Manual 
  & Manual 
  &  Automated \\
No API knowledge required 
  & $\times$ 
  & $\times$ 
  & \checkmark \\
Spec.\ defect handling 
  & $\times$ 
  & $\times$ 
  &  Detect + repair \\
Tool-set reduction 
  & $\times$ 
  & $\times$ 
  &  Filter + group \\
Evaluation scope 
  & 1 domain 
  & $\times$ 
  &  77 APIs \\
\bottomrule
\end{tabular}
\end{table}

Any specification-driven approach depends on specification quality, 
which empirical research has shown to be 
uneven~\cite{atlidakis2019restler, golmohammadi2023testing}. 
Incomplete security declarations, undocumented parameters, and 
specification--implementation mismatches are documented across 
public OpenAPI corpora~\cite{uddin2015api}. The same defect classes 
impede MCP server generation, yet no prior work has addressed their 
detection and repair in this context. \textsc{AutoMCP} bridges this 
gap: it requires no API-specific developer knowledge, extracts all 
configuration from the specification, repairs its defects 
automatically, and applies empirically grounded tool-set 
transformations, evaluated across 77~real-world APIs.

%% file: Sections/ResearchMethod.tex
\section{Study Design}
\label{sec:study_design}

\subsection{Research Questions and Study Overview}
\label{sec:rq_overview}

In this paper, we investigate four research questions (RQs) designed to provide a structured examination of MCP server design and automation. Specifically, we seek to answer the following questions:

\begin{itemize}
\item \textbf{RQ1:} To what extent do MCP servers rely on vendor REST APIs, and what integration strategies and configuration practices do they adopt?

\item \textbf{RQ2:} How are vendor REST API operations translated into MCP tools, in terms of exposure, omission, and mapping patterns?

\item \textbf{RQ3:} To what extent can REST-backed MCP servers be automatically generated from OpenAPI specifications, and what factors prevent faithful tool generation?

\item \textbf{RQ4:} How effective are automated specification repair and tool-set transformations in addressing generation failures and tool-set explosion?
\end{itemize}

\begin{figure}[h]
\centering
\includegraphics[width=\linewidth]{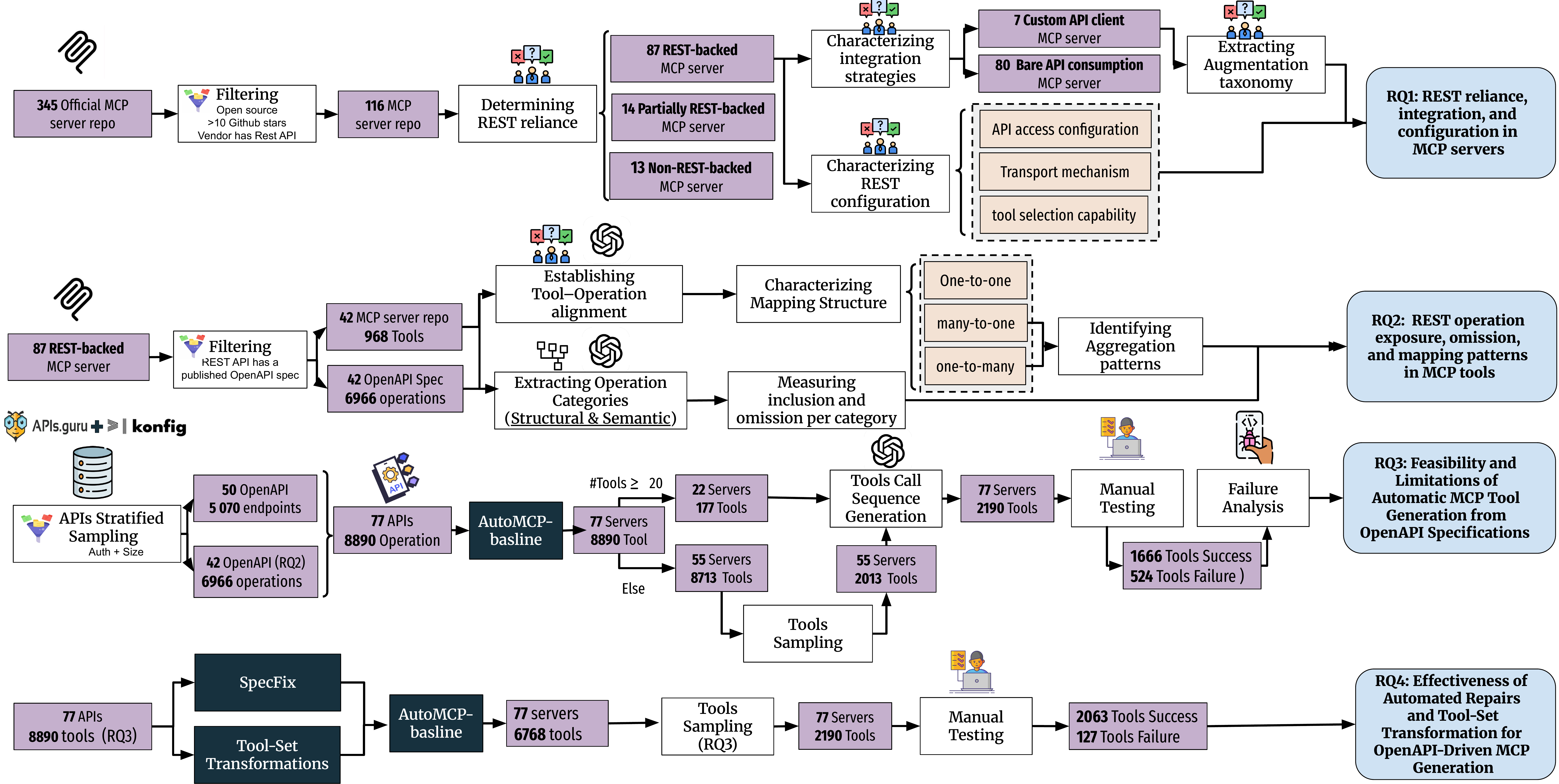}
\caption{Approach overview and addressed RQs.}
\label{fig:approach}
\end{figure}

Figure~\ref{fig:approach} presents an overview of our study design, outlining the multi-stage analysis conducted to answer the four research questions.  (\textbf{RQ1}) characterizes how MCP servers rely on and integrate REST APIs in practice. (\textbf{RQ2}) builds on this by examining how API functionality is exposed at the tool level. (\textbf{RQ3}) then evaluates the extent to which such servers can be generated automatically from OpenAPI specifications. Finally, (\textbf{RQ4}) addresses the limitations identified in RQ3 through specification repair and tool-set transformations. 

\subsection{Data Collection}
\label{sec:dataset}

Our study relies on three related datasets constructed to support the analyses for RQ1 through RQ4.

For \textbf{RQ1}, we started from the official MCP server catalogue maintained by Anthropic as of July~31,~2025 \cite{mcp_servers_repo}, which listed 345 publicly accessible MCP servers. After removing entries that did not correspond to actual source-code repositories (e.g., documentation stubs, placeholder listings, or non-functional demonstrations), 279 repositories remained. We then restricted the dataset to the 165 repositories whose vendors publish a publicly documented REST API, since these systems are capable of expressing REST functionality through MCP tools. To focus on mature and non-trivial implementations, we excluded repositories with fewer than 10 GitHub stars. Prior empirical software engineering studies commonly use star counts as indicators of project maturity and community engagement~\cite{valiev2018ecosystem,borges2018understanding}. This filtering step yielded a final dataset of \textbf{116 MCP server repositories} for RQ1.

For \textbf{RQ2}, we required both an MCP server implementation and a machine-readable API specification. We therefore selected the subset of the RQ1 corpus for which the vendor publishes an official OpenAPI  specification. This process resulted in \textbf{42 MCP servers} with corresponding OpenAPI contracts, forming the dataset used for all mapping analyses.

\textbf{RQ3} and \textbf{RQ4} evaluate \textsc{AutoMCP}, which requires a broader and more diverse set of API specifications than the MCP-focused dataset used in RQ1--RQ2. To construct this extended dataset, we combined the 42 APIs from RQ2 with a stratified random sample of 50 OpenAPI~3.x specifications drawn from two large public API collections: \emph{APIs.guru} and the \emph{Konfig} dataset. After deduplication and removal of syntactically invalid specifications, these sources provided 3,784 unique machine-readable OpenAPI contracts. 

We performed stratified sampling based solely on metadata that can be computed automatically: authentication scheme (none, API key, basic/bearer, OAuth~2.0) and specification size. Size buckets were defined empirically using endpoint-count percentiles: small ($\leq$20 operations), medium (21--100), and large (>100). Random sampling without replacement continued until each \{authentication, size\} cell was represented and the target sample size of 50 specifications was reached. To support reconstruction-fidelity evaluation, we augmented the stratified sample with the 42 APIs from RQ2. After deduplication, the resulting evaluation corpus for RQ3--RQ4 consisted of \textbf{80 unique OpenAPI specifications}.

\section{RQ-Specific Methodology}
\label{sec:rq_methodology}
\subsection{RQ1: To what extent do MCP servers rely on vendor REST APIs, and what REST integration
strategies and configuration practices do they adopt?}
\label{sec:rq1-approach}

\vspace{0.1cm}

\textbf{Motivation.} As MCP is promoted as an open standard for connecting LLM applications to external tools and data sources via MCP servers, a growing set of “official integrations” is being published as reference/maintained servers in public repositories and registries \cite{mcp_servers_repo}. In practice, many of these official servers appear to target the same underlying vendor capabilities already available via their public REST APIs. Yet, we lack empirical evidence on whether official MCP servers are predominantly REST-backed wrappers (and whether REST is used uniformly across tools or only selectively) or whether other backends dominate. Moreover, even among REST-backed servers, it is unclear how REST capabilities are integrated into tool implementations, specifically, whether tools primarily perform bare API access or instead incorporate additional adaptation/augmentation logic beyond routine API access, and which configuration practices (e.g., authentication and transport setup, tool filtering) are adopted in practice.

\vspace{0.2cm}
\textbf{Approach.} 
\label{sec1}
To answer RQ1, we analyze the 116 MCP servers described in Section~\ref{sec:dataset}. We conducted a two-stage manual analysis to (i) determine the extent to which each server relies on the vendor REST API and (ii) characterize, for REST-backed servers, the concrete REST integration strategies and configuration practices used in implementation.

\vspace{0.2cm}

\underline{\textit{Stage 1: Determining REST reliance.}} Two authors independently inspected each repository using a fixed trace procedure, with a third author involved in the adjudication. First, we identified all tools exposed by the server by locating tool registration and handler dispatch points. Second, for each tool, we traced the implementation from the handler entry point to the first external-service interaction by following function calls and imported modules. Third, we determined whether the interaction targets the vendor’s REST API by inspecting request destinations (e.g., base URLs/hosts), API client initialization code, and dependencies indicative of vendor SDKs or HTTP clients. Based on these tool-level traces, each repository was labeled as \emph{REST-backed} (all tools invoke the vendor REST API), \emph{partially REST-backed} (only a subset of tools invoke the vendor REST API), or \emph{non-REST-backed} (no tool invokes the vendor REST API). Inter-rater reliability on the pre-consensus labels was assessed using Cohen’s kappa for nominal categories ($\kappa=0.87$). Disagreements were resolved through an evidence-based adjudication workflow: the two raters exchanged written justifications grounded in concrete code artifacts (file paths, functions, imports, and call sites) and then held a reconciliation meeting.

\vspace{0.2cm}

\underline{\textit{Stage 2: Characterizing REST configuration and integration strategies.}} 
For repositories labeled \emph{REST-backed}, we analyzed two complementary aspects: (i) configuration practices governing authentication, transport, and tool exposure, and (ii) integration strategies describing how REST API calls are incorporated into tool implementations. We conducted an exhaustive manual review using a structured coding protocol and a shared codebook. From the 87 REST-backed servers identified in Stage~1, we randomly selected 25 for a Stage~2 pilot and had all three authors code them independently. We compared pilot results to refine the protocol’s operational definitions and decision rules. Coders inspected each repository in a fixed order: (1) tool handler implementations, (2) end-to-end call paths from handlers to external interactions (including client construction and request execution), and (3) configuration artifacts and initialization code. We then split the remaining 62 repositories across authors ($\approx 20$ each), holding periodic calibration meetings and escalating ambiguous cases for consensus.

\textbf{(i) Configuration practices.} Across all \emph{REST-backed} repositories, we first documented configuration practices relevant to API access and execution context by examining configuration files, environment variable templates, and initialization code. Concretely, we coded (i) \emph{API access configuration} (credential provisioning/authentication setup and how credentials are supplied), (ii) \emph{transport mechanism support} (STDIO, HTTP, SSE, or combinations), and (iii) \emph{tool selection/filtering capability} (whether subsets of tools can be enabled, disabled, or filtered via configuration or registration logic). Reliability for these repository-level labels was assessed on pre-consensus pilot labels using Krippendorff’s $\alpha$ ($\alpha_{\text{auth-config}}=0.94$, $\alpha_{\text{transport}}=0.96$, $\alpha_{\text{tool-filtering}}=0.90$), after which disagreements were reconciled using code-grounded evidence.

\textbf{(ii) Integration strategies (bare vs.\ custom).}
We then characterized REST integration strategies at the tool level by distinguishing whether a tool’s implementation performs only API access or API access plus additional custom logic. A tool was labeled \emph{bare API consumption} when its logic primarily (i) maps tool parameters to an external operation, (ii) issues the request/command, and (iii) returns the external response with at most routine client concerns (e.g., authentication setup, request formatting, pagination traversal, and error handling). In contrast, a tool was labeled \emph{custom API client} when it routes external interactions through a repository-defined client/wrapper that performs \emph{non-trivial integration logic beyond routine API access}, i.e., intermediate processing that materially alters how vendor capabilities are invoked or combined beyond standard request/response handling. Reliability for the \emph{bare} vs.\ \emph{custom} tool label was assessed on pre-consensus pilot labels (Krippendorff’s $\alpha_{\text{bare/custom}}=0.85$), followed by reconciliation among the three authors using concrete call sites, imports, and execution traces as evidence. Finally, we lifted tool-level labels to the server level using an existence rule: a server was labeled \emph{custom API client} if it contained at least one tool labeled \emph{custom API client}; otherwise, it was labeled \emph{bare API consumption}.

\textit{\textbf{(ii-a) Invocation mechanisms for bare API consumption MCP servers.}}
For MCP servers labeled \emph{bare API consumption}, we further coded the concrete invocation type used to access vendor capabilities as \emph{direct HTTP} (e.g., \texttt{fetch}, \texttt{requests}), \emph{SDK-based} (vendor/third-party SDKs), \emph{specification-driven} (e.g., OpenAPI-generated clients), or \emph{CLI-based} (invoking vendor CLIs via subprocess execution). Reliability for this multi-class label was assessed on pre-consensus pilot labels (Krippendorff’s $\alpha_{\text{bare-subtype}}=0.83$), followed by consensus reconciliation grounded in concrete call sites and dependency evidence.

\textit{\textbf{(ii-b) Augmentation taxonomy for custom API client MCP servers.}}
Finally, tools labeled \emph{custom API client} were flagged for additional characterization to capture augmentation logic beyond routine API access. While coding the non-pilot repositories, coders marked every \emph{custom API client} tool for follow-up. After primary coding of all REST-backed repositories, we pooled the flagged tools and derived an augmentation taxonomy via open coding. For each flagged tool, all three authors independently traced (i) the repository-defined client/wrapper and (ii) its call sites in tool handlers to identify any intermediate processing between tool invocation and external interactions, and assigned descriptive labels for the observed augmentation behaviors. We then consolidated these labels into a provisional taxonomy by merging semantically equivalent labels and splitting overly broad labels. Using this provisional taxonomy, we mapped each author’s labels to taxonomy categories and computed pre-consensus agreement on category assignments across all flagged tools (Krippendorff’s $\alpha_{\text{custom\_category}}=0.81$). We resolved disagreements in code-grounded reconciliation meetings and refined category names and definitions to produce the final augmentation taxonomy.

\vspace{0.3cm}

\subsection{RQ2: How do MCP servers translate vendor REST API operations into tools, specifically, what proportion of available operations is exposed, which classes of operations are omitted, and what mapping patterns are used to group operations into tools?}
\label{subsec:rq2-approach}

\vspace{0.1cm}

\textbf{Motivation.} RQ1 shows that the majority of official MCP servers act as REST API wrappers, but this observation alone does not characterize the practical functionality that these servers make available to MCP clients. Prior work has shown that APIs can be difficult to use at scale, and that the level of abstraction and the structuring of operations are central factors influencing API usability and developer experience \cite{Robillard2009,Murphy2018}. At the same time, studies on API operation design show that APIs often move beyond a strictly CRUD-based representation of underlying resources, instead offering operations that are organized around specific tasks \cite{Singjai2021}. Moreover, security guidance emphasizes minimizing exposed capabilities and avoiding unnecessarily privileged or sensitive endpoints, since API interfaces can expand an application’s attack surface if not carefully scoped \cite{OWASP2023}. Together, these strands suggest that MCP servers should not be assumed to expose vendor APIs verbatim: instead, they likely implement systematic selection (what is exposed vs.\ omitted) and abstraction (how endpoints are grouped into tools). RQ2 therefore, examines the operation-to-tool mapping to quantify exposure and omission across operation/feature classes and to identify recurring one-to-one,  many-to-one and one-to-many mapping structures.

\vspace{0.2cm}

\textbf{Approach.}
To answer RQ2, we analyze how MCP servers map vendor REST operations to tools. We restrict this analysis to \emph{REST-backed} and \emph{Partially-REST-backed} MCP servers, excluding non-REST-backed servers identified in RQ1, and focus on the subset of 42 servers with corresponding OpenAPI specifications described in Section~\ref{sec:dataset}. 
On this subset, we perform a three-stage analysis to (i) establish a mapping between MCP tools and REST operations, (ii) quantify operation exposure and omission across operation classes (e.g., read/create/update/delete) and endpoint categories (e.g., authentication/security and analytics), and (iii) characterize mapping structures, including one-to-one mappings, many-to-one aggregation, and one-to-many reuse.


\vspace{0.2cm}

\underline{\textit{Stage 1: Establishing tool--operation alignment.}}
We establish an alignment between MCP tools and vendor REST operations by combining manual mapping on a gold-standard subset with an LLM-assisted semantic matching. We reuse the artifacts extracted during RQ1: for each MCP server we have the set of exposed tools (from tool registration/handler definitions) and, for each tool, its name, natural-language description, and parameter schema when available, together with evidence about how the tool invokes the vendor (e.g., HTTP vs.\ SDK vs.\ CLI, and any explicit method/path or URL templates observable at HTTP call sites). In parallel, we parse each vendor’s OpenAPI specification to extract the set of REST operations, treating an operation as an HTTP method bound to a path template (e.g., \texttt{GET /users/{id}}), and record operation metadata (e.g., \texttt{operationId}, summary/description, tags, and request schema when present).

To construct a gold-standard alignment, we randomly sampled 10 MCP servers from the 42-server corpus using a fixed seed. This subset contains 298 MCP tools and 857 REST operations. For each tool in the gold set, three authors independently mapped the tool to one or multiple REST operations. Because many tools invoke vendor functionality through SDKs or CLIs that abstract away the underlying REST endpoints, we treat tool--operation alignment as a semantic mapping problem and base the mapping primarily on textual and structural metadata (tool name/description and parameter schema; OpenAPI operation summary/description, \texttt{operationId}, HTTP method, and path template). When the tool's HTTP call sites explicitly reveal endpoint information (e.g., method/path fragments or URL templates), we used this evidence to rule out non-matching OpenAPI operations and prioritize candidates consistent with the observed endpoint.

We assessed agreement at two levels on the pre-consensus annotations. First, we measured agreement on mapping cardinality, whether annotators selected a single operation versus multiple operations for a given tool, using a multi-rater nominal agreement statistic (Krippendorff's $\alpha_{\text{card}}=0.82$). Second, we measured agreement on mapping identity by comparing, for each tool, the sets of operations selected by different annotators, computing a set-overlap similarity score per tool, and macro-averaging this score across tools and annotator pairs ($\overline{\text{SetSim}}=0.89$). Disagreements were then resolved through evidence-based reconciliation until consensus was reached.

To scale beyond the gold-standard alignment, we employ an LLM (GPT-4.1) as a semantic labeler, following established practices in LLM-assisted annotation and coding workflows where a human-derived protocol is operationalized as a structured prompt and validated against gold labels before being applied at scale \cite{dunivin2025scaling}. Concretely, we translate our manual mapping guidelines into an instruction prompt. To reduce spurious matches and improve determinism, the prompt requires the model to choose only from the operations provided in the prompt and to return the selected operation identifiers in a structured format (accompanied by a brief justification to support auditing).

We evaluate the LLM against the gold alignment at the tool level. For each tool, we compare the predicted set of one or more operations to the gold set and report macro-averaged per-tool precision, recall, and F$_1$. We additionally report micro-averaged precision/recall/F$_1$ over all tool--operation decisions, as well as exact-match accuracy and cardinality accuracy to capture whether the model predicts the correct operation set and set size. On the gold set, the LLM achieves high performance (per-tool precision is 0.90, recall is 0.84, F$_1$ is 0.87; micro-F$_1$  is 0.89; exact-match is 0.78 and cardinality accuracy is 0.82). After evaluation, we apply the same prompt and matching procedure unchanged to the remaining servers to generate tool--operation alignments at scale.

\vspace{0.2cm}

\underline{\textit{Stage 2: Quantifying operations coverage and omission in MCP tools}}
\vspace{0.05cm}

Using the alignment results of Stage~1, we quantify the extent to which vendor REST APIs are exposed through MCP servers and how exposure varies across operation classes.

\paragraph{A.\ Overall operation coverage.}
For each API $s$, let $O_s$ denote the set of REST operations defined in its OpenAPI specification and let $M_s \subseteq O_s$ denote the subset mapped to at least one MCP tool. We define per-API coverage as:
\begin{equation}
\mathrm{Cov}_s\left(\ast\right)=\frac{|M_s|}{|O_s|}.
\label{eq:cov_overall}
\end{equation}

We analyze the distribution of $\mathrm{Cov}_s\left(\ast\right)$ across APIs by reporting its median and mean, and examine how coverage varies as a function of API size.

\paragraph{B.\ Coverage and omission by operation category.}
We next examine how coverage varies across different types of API operations. Using the tool--operation alignment from Stage~1, we group operations by structural and semantic categories derived from their OpenAPI specifications and compute coverage/omission within each category. Accordingly, we label every REST operation with two classes of features: \emph{structural} and \emph{semantic}.

\textbf{Structural features}, illstrated in \autoref{tab:rq2_struct_features}, capture the interface role and structural complexity of an operation and are computed deterministically from the OpenAPI specification, following prior work on operation-level API analysis \cite{Serbout2024APIsticAL,bogner2020,haupt2017,serbout2}. We derive \texttt{is\_read\_only}, \texttt{is\_mutating}, and \texttt{is\_destructive} from the HTTP method (GET vs.\ POST/PUT/PATCH vs.\ DELETE). We derive \texttt{has\_parameters} and \texttt{has\_request\_body} from the presence of declared parameters (path/query) and a \texttt{requestBody} object, respectively. We label \texttt{needs\_authentication} when the effective operation specifies OpenAPI security requirements and \texttt{is\_deprecated} when the operation is marked as deprecated. Finally, we detect \texttt{has\_pagination} when the operation declares common pagination controls as query parameters (e.g., \texttt{page}, \texttt{offset}, and \texttt{cursor}). To capture navigational complexity, we compute \texttt{path depth} as the number of segments in the normalized OpenAPI path template. Depth-/path-length notions have been used in prior REST/OpenAPI structural analyses to characterize how ``deep'' an API's resource structure becomes (e.g., via longest-path measures)\cite{haupt2018api}. Because APIs differ substantially in size and path conventions, we adopt an API-relative threshold to identify unusually deep operations: an operation is labeled \texttt{deep-path} if its depth exceeds $\max\{Q_{80}, 5\}$, where $Q_{80}$ is the API-specific 80th percentile of path depth. This follows the general idea of relative thresholds that define cutoffs from the distribution of values rather than a single global constant, while the minimum cutoff of $5$ prevents uniformly shallow APIs from labeling ordinary endpoints as deep~\cite{oliveira2014extracting}.

\textbf{Semantic features} capture higher-level functional groupings that are not directly encoded by structural OpenAPI fields. We derive semantic categories by clustering OpenAPI tags into higher-level groups. We extract all tag strings from the specifications and normalize them deterministically (lowercasing and trimming whitespace, standardizing separators, and removing version tokens/stopwords). We embed each normalized tag using a pretrained sentence encoder (Sentence-Transformers \texttt{all-MiniLM-L6-v2} \cite{all-minilm-l6-v2}). To reduce sensitivity to any single run, we use a bootstrap consensus procedure with $B=30$ rounds: in each round, we resample tags with replacement and fit BERTopic \cite{grootendorst2022bertopic} using the precomputed embeddings (UMAP \cite{mcinnes2018umap} + HDBSCAN \cite{campello2013density} internally), yielding a clustering of the sampled tags (including an outlier label). Across rounds, we accumulate a co-association score for each tag pair based on how often the two tags are assigned to the same cluster when they co-appear; we then run HDBSCAN on the resulting consensus distance matrix to obtain final tag groups. We treat outliers and low-stability clusters as unreliable and exclude them from category-level reporting. Each REST operation inherits the semantic category (or categories) of its OpenAPI tag(s); semantic categories are non-mutually-exclusive, and operations may contribute to multiple categories when they carry multiple tags. To obtain human-interpretable category names, Three authors independently reviewed each cluster’s most frequent tags and representative operations, proposed a concise label, and reconciled naming disagreements in a meeting without modifying cluster membership. Finally, to focus on broadly recurring categories, we retain only semantic categories that appear in at least 25\% of APIs in our corpus.

\begin{table}[t]
\centering
\small
\setlength{\tabcolsep}{6pt}
\begin{tabularx}{\linewidth}{lX}
\toprule
\textbf{Feature} & \textbf{Definition} \\
\midrule
\texttt{is\_read\_only} &
Operation uses HTTP \texttt{GET}. \\
\texttt{is\_mutating} &
Operation uses \texttt{POST}, \texttt{PUT}, or \texttt{PATCH}. \\
\texttt{is\_destructive} &
Operation uses HTTP \texttt{DELETE}. \\
\texttt{has\_parameters} &
Operation declares at least one path or query parameter. \\
\texttt{has\_request\_body} &
Operation includes a request body. \\
\texttt{has\_pagination} &
Operation exposes pagination controls (e.g., \texttt{page}, \texttt{offset}, \texttt{cursor}). \\
\texttt{needs\_authentication} &
Operation is protected by OpenAPI security requirements. \\
\texttt{is\_deprecated} &
Operation is marked as deprecated in the specification. \\
\texttt{deep\_path} &
Operation path depth exceeds an API-specific threshold. \\
\bottomrule
\end{tabularx}
\caption{Structural operation-level features used to quantify coverage and omission.}
\label{tab:rq2_struct_features}
\end{table}

\textbf{Category-Level Metrics}
For a given category $c$ (structural or semantic) and API $s$, let $O_{s,c}\subseteq O_s$ be the operations exhibiting $c$, and let $M_{s,c}=M_s\cap O_{s,c}$ be those mapped to at least one tool. Let $S_c \subseteq S$ denote the set of APIs for which category $c$ is present, i.e., those with $|O_{s,c}|>0$. Category-level coverage and omission are:
\begin{equation}
\mathrm{Cov}_s\left(c\right)=\frac{|M_{s,c}|}{|O_{s,c}|}\quad \left(|O_{s,c}|>0\right),
\label{eq:cov_cat}
\end{equation}
\begin{equation}
\mathrm{Omit}_s\left(c\right)=1-\mathrm{Cov}_s\left(c\right).
\label{eq:omit_cat}
\end{equation}

We aggregate coverage across APIs using both macro and micro averages. The macro average captures per-API behavior independent of size:
\begin{equation}
\mathrm{Cov}^{\mathrm{macro}}\left(c\right)=\frac{1}{|S_c|}\,\mathop{\Sigma}\limits_{s\in S_c}\,\mathrm{Cov}_s\left(c\right),
\label{eq:cov_macro_cat}
\end{equation}
while the micro average reflects operation-weighted coverage:
\begin{equation}
\mathrm{Cov}^{\mathrm{micro}}\left(c\right)=\frac{\mathop{\Sigma}\limits_{s\in S_c}|M_{s,c}|}{\mathop{\Sigma}\limits_{s\in S_c}|O_{s,c}|},
\label{eq:cov_micro_cat}
\end{equation}

To quantify whether a category is \emph{preferred} for exposure relative to an API's baseline, we compute per-API lift:
\begin{equation}
\mathrm{Lift}_s\left(c\right)=\frac{\mathrm{Cov}_s\left(c\right)}{\mathrm{Cov}_s\left(\ast\right)},
\label{eq:lift}
\end{equation}
where $\mathrm{Lift}_s\left(c\right)>1$ indicates that operations in category $c$ are mapped at a higher rate than the API's overall coverage. We summarize lift using macro aggregation:
\begin{equation}
\mathrm{Lift}^{\mathrm{macro}}\left(c\right)=\frac{1}{|S_c|}\,\mathop{\Sigma}\limits_{s\in S_c}\,\mathrm{Lift}_s\left(c\right),
\label{eq:lift_macro}
\end{equation}

\vspace{0.2cm}

\underline{\textit{Stage 3: Characterizing grouping patterns in tool--operation mappings.}}
\vspace{0.2cm}

\textbf{Mapping Structure}
Using the tool--operation alignment from Stage~1, we first quantify mapping structure from both perspectives. From the tool perspective, we classify a tool as \emph{one-to-one} if it aligns to exactly one REST operation, and as \emph{many-to-one} if it aligns to multiple operations. From the operation perspective, we classify an operation as \emph{one-to-one} if it aligns to exactly one tool, and as \emph{one-to-many} if it aligns to multiple tools.

\textbf{Aggregation Pattern Identification}
We then investigate recurring grouping rationales for many-to-one aggregation and one-to-many reuse through manual labeling. We treat each many-to-one tool as one analysis instance and each one-to-many operation as one analysis instance. For each instance, all three authors independently assigned a free-form pattern label capturing the perceived rationale for the grouping, along with a short justification grounded in evidence from (i) tool metadata, (ii) aligned operation metadata, and, when necessary, (iii) repository code evidence indicating how multiple operations are orchestrated or reused.

To quantify agreement before reconciliation, we computed inter-annotator reliability over the pre-consensus labels for many-to-one and one-to-many instances separately (Krippendorff’s $\alpha$ for nominal labels; $\alpha_{\text{m2o}}=0.91$ and $\alpha_{\text{o2m}}=0.94$). We operationalized label agreement as assigning the same instance to the same pattern category; disagreements were not reconciled prior to computing $\alpha$. After agreement measurement, we held synthesis meetings to resolve disagreements using the recorded justifications and underlying evidence, and to consolidate synonymous labels and standardize final pattern names. Finally, to focus on recurring design choices rather than API-specific idiosyncrasies, we retain only patterns that appear in at least two distinct APIs; patterns observed in exactly one API are excluded.

\vspace{0.3cm}

\subsection{RQ3: To what extent can REST-backed MCP servers be automatically generated from OpenAPI specifications, and what factors prevent faithful tool generation in practice?}
\label{subsec:rq3-approach}

\vspace{0.1cm}

\textbf{Motivation.} Building on RQ1 and RQ2, our empirical characterization suggests that specification-driven automation should be feasible in principle: most official MCP servers are REST-backed, predominantly implement bare API consumption, and a substantial portion of tool–API relationships follow direct one-to-one mappings. However, OpenAPI-based MCP generation appears rarely in practice ($\approx4.5\%$ of servers in our corpus), despite OpenAPI being explicitly intended as a machine-readable, language-agnostic interface description for HTTP APIs that enables consumers to discover and understand service capabilities without access to source code \cite{kim2023enhancing}. Moreover, OpenAPI is widely used as a foundation for downstream automation across the API lifecycle, including documentation, client generation, and testing \cite{ed2018automatic}. These observations motivate RQ3, which empirically evaluates, under a fixed baseline translation, the extent to which real-world OpenAPI specifications suffice for generating REST-backed MCP servers and tools, and identifies the recurring factors that prevent faithful tool generation in practice.

\vspace{0.2cm}

\textbf{Approach.}
To assess OpenAPI-driven MCP generation in a controlled, reproducible setting, we developed AutoMCP-baseline (Section~\ref{sec:automcp}), a baseline generator that automatically translates OpenAPI specifications into REST-backed MCP servers. AutoMCP-baseline implements a fixed and transparent operation-to-tool translation: each OpenAPI operation (HTTP method bound to a path template) is converted into a distinct MCP tool, and tool schemas and metadata are derived solely from the specification. We adopt this prototype because our objective is not to benchmark community-developed OpenAPI-to-MCP tools, but to measure what OpenAPI alone enables under a uniform baseline translation. Existing open-source solutions are often provided as libraries/frameworks that require user-authored integration code and configuration choices, which would introduce uncontrolled variation across APIs and confound corpus-scale measurement. AutoMCP-baseline therefore serves as a controlled measurement instrument with standardized logging, enabling consistent evaluation and failure analysis across our specification corpus.

We conduct this study on a corpus of $80$ real-world OpenAPI specifications described in Section~\ref{sec:dataset}. For each specification, AutoMCP-baseline applies a baseline “operation-to-tool” translation: every documented operation (HTTP method bound to a path template) is translated into a distinct MCP tool, with operation inputs consolidated into a single tool schema and operation metadata used to populate tool names and descriptions.

Because executing every generated tool across all APIs is infeasible, we construct an evaluation set of executable operations using a two-phase sampling strategy inspired by resource-centric REST testing and structural coverage principles \cite{atlidakis2019restler,corradini2021restats}. In the first phase, for APIs with at most 20 operations, we attempt to execute all operations. In the second phase, for APIs with more than 20 operations, we stratify operations by resource group, defined by the first semantic path segment (e.g., \texttt{/users}, \texttt{/repos}), and select a structurally diverse subset within each group. Specifically, we score each operation $o$ by its contribution to uncovered diversity dimensions:

\begin{equation}
s\text{(}o\text{)}=\Ind{verb}\tplus\Ind{auth}\tplus\Ind{params}
\label{eq:rq3_diversity_score}
\end{equation}

where an operation earns one point if it introduces a previously unseen HTTP verb within the group, one point if it introduces a new authentication scheme (from OpenAPI security requirements), and one point if it introduces a new parameter modality (path/query/header/body). We then apply a greedy pass that repeatedly selects the highest-scoring remaining operations until each diversity axis observed in the group is covered at least once. Across both phases, endpoints that require paid subscriptions, geo-restricted credentials, or unavailable accounts are excluded. This sampling procedure yields a final evaluation set of $2,190$ executable operations (i.e., generated MCP tools) across the corpus.

We execute sampled tools using Claude  as the MCP client, explicitly selecting tools from the loaded manifest to avoid confounding results with free-form tool discovery. Because many tools exhibit logical dependencies (e.g., a resource must be created before it can be updated or deleted), execution order is non-trivial. We therefore first use GPT-4.1 to propose an initial dependency-respecting sequence over the sampled tools based on tool names and descriptions, following recent practice for planning API call sequences \cite{le2024kat,kim2024multiagent_rest_api_testing}. We then partition the resulting sequence into dependency blocks that group tools sharing state (e.g., create→read→update→delete for the same resource family) and assign these blocks to one of three authors to balance workload while ensuring state-dependent tools are executed by the same author. After the assignment, each author reviews the plan for their block(s) and revises it when necessary (e.g., inserting missing prerequisites or correcting ordering), logging all edits as part of the evaluation trace.

For each assigned tool, the author issues a minimal, template-based prompt intended to exercise the MCP server rather than assess LLM reasoning. Prompts follow a consistent structure: they explicitly name the tool to invoke, provide concrete, valid input values (reusing identifiers produced earlier in the same dependency block when required), and state an observable expectation when applicable (e.g., “create X, then verify via list/get”). A tool execution is considered successful if (1) the tool appears in the manifest and loads without schema/configuration errors, (2) the invocation receives a successful HTTP response from the upstream API, and (3) the intended effect is observed, verified either by inspecting returned data (for read operations) or via a follow-up state query within the same block (for update/write/delete operations). For every tool, we record the prompt, concrete inputs, HTTP status codes, and response payloads, and a success/failure label. To assess labeling consistency, a 10\% random subset of executed tools is independently re-evaluated by a second author using the same block plan and success criteria; Cohen's $\kappa$ on the re-evaluated subset was $0.97$. Disagreements were resolved by jointly reviewing the recorded prompts and HTTP transcripts; remaining cases were adjudicated by a third author, and the adjudicated label was used as final.

We analyze failed executions to identify recurring obstacles to faithful OpenAPI-driven MCP generation. For each failed tool invocation, three authors independently inspected AutoMCP-baseline server logs, MCP client traces, and HTTP request/response transcripts and assigned a provisional root-cause label capturing the primary reason for failure. We quantified inter-annotator agreement on these pre-consensus labels using Krippendorff's $\alpha$ for nominal data, where agreement is defined by whether annotators assigned the same failure to the same root-cause category ($\alpha=0.93$). The authors then held a synthesis meeting to reconcile disagreements and consolidate synonymous labels into a shared taxonomy with explicit definitions. Using this finalized taxonomy, each failure was assigned a single final root-cause label.

\vspace{0.3cm}
\subsection{RQ4: How effective are automated OpenAPI repairs and tool-set transformations (filtering and grouping) at addressing the failures and tool-set explosion observed in OpenAPI-driven MCP generation?}
\label{subsec:rq4-approach}

\vspace{0.2cm}

\textbf{Motivation.}
RQ3 demonstrates that OpenAPI-driven generation can fail to produce faithful MCP servers in practice, and that even when generation succeeds, the resulting tool interfaces can be impractically large for real MCP clients. These findings indicate two distinct barriers to automation. First, generation failures often arise when OpenAPI specifications are incomplete or incorrect relative to vendor behavior, preventing valid tool schemas or reliable request execution. Second, baseline operation-to-tool translation tends to produce a tool per operation, which can yield hundreds or thousands of tools for large APIs, exceeding practical client limits and undermining usability. At the same time, RQ2 provides empirical evidence that real MCP servers do not expose vendor APIs verbatim: they routinely omit certain categories of operations and often apply many-to-one grouping patterns to present higher-level tools. Together, these observations motivate RQ4: we investigate whether the limitations revealed in RQ3 can be addressed automatically by repairing OpenAPI specifications using vendor documentation cues and by transforming the generated tool set using empirically grounded filtering and grouping strategies.

\vspace{0.2cm}
\textbf{Approach.}
RQ4 assesses whether the correctness and usability of OpenAPI-driven MCP servers can be improved automatically through two complementary strategies: (i)~detecting and repairing specification defects using \textsc{SpecFix} (Section~\ref{sec:specfix}), and (ii)~reducing tool-set size through category-based filtering and deterministic grouping (Section~\ref{sec:transformations}). Implementation details of both components are provided in the respective framework sections; here we describe the evaluation methodology.  We use the same corpus of OpenAPI specifications as described in Section~\ref{sec:dataset}, as well as the same baseline generation and execution harness introduced in RQ3, including the sampled operations per API, the prompt templates, and the recorded execution transcripts, and proceeds in two corresponding stages.

\vspace{0.2cm}

\underline{\textit{Stage 1: Specification defect detection and repair.}}

\vspace{0.1cm}

\textsc{SpecFix} (Section~\ref{sec:specfix}) operates in two modes on each specification. In \emph{detection} mode, it analyses the OpenAPI document together with vendor documentation cues and emits a structured issue report identifying suspected defects and the operations they affect. In \emph{repair} mode, it consumes the issue report and synthesises a patch for each detected defect that minimally amends the OpenAPI document while preserving its intended semantics. The evaluation targets defect categories A--C (Table~\ref{tab:rq4_specfix}), which are specification-level defects identifiable through static analysis of the OpenAPI document and its associated documentation. Categories D and E are excluded because they manifest only at runtime and require dynamic request--response analysis to detect reliably~\cite{atlidakis2019restler,golmohammadi2023testing}.

We evaluate detection quality on the 16 APIs whose RQ3 failure traces were attributed to specification-level defect categories A--C (Section~\ref{sec:specfix}), encompassing 20 defect instances. Three authors independently assess whether each reported defect corresponds to a real inconsistency or omission in the OpenAPI document, given the referenced documentation evidence and the baseline failure traces from RQ3. Inter-annotator agreement on the pre-consensus judgements is quantified using Krippendorff's $\alpha$ for nominal labels ($\alpha = 1.0$).

We evaluate repair quality on the same 16 APIs by checking both validity and minimality of the produced patches. At the specification level, we verify that patched documents remain parseable and that modifications are localised to the reported defect context. At the semantic level, three authors independently assess whether each patch plausibly resolves the reported issue while preserving intended behaviour, based on the original OpenAPI fragment, the documentation evidence referenced in the issue report, and the patch diff. Agreement is quantified using Krippendorff's $\alpha$ ($\alpha = 0.93$), followed by reconciliation.

We then measure end-to-end recovery by rerunning the RQ3 baseline pipeline on the full corpus of 77 APIs after applying \textsc{SpecFix} repairs, and comparing generation and execution outcomes to the original baseline. At the server level, we evaluate whether the regenerated MCP manifest loads successfully. At the tool level, we re-execute, for each API, the same sampled operation set used in RQ3, preserving tool intents and dependency ordering. Tool executions are judged using the same success criteria as in RQ3, and we report improvements in execution success rate, reductions in specification-related failure categories, and regressions in which previously successful invocations fail after repair. To verify labelling consistency, a second author independently re-assessed a random 10\% sample of the repaired-tool execution outcomes using the same success criteria; inter-annotator agreement was perfect (Krippendorff's $\alpha = 1.0$).

\vspace{0.2cm}

\underline{\textit{Stage 2: Tool-set reduction via filtering and grouping.}}

\vspace{0.1cm}

The second stage addresses tool-set explosion by transforming the baseline one-operation-per-tool output into a smaller, more usable set of tools. The transformation pipeline—category-based filtering followed by deterministic grouping—is described in Section~\ref{sec:transformations}; here we detail the evaluation methodology.

\emph{Category labelling and validation.}
Filtering and grouping both require assigning structural and semantic category labels to each OpenAPI operation. Structural labels are extracted directly from the specification. For semantic labels, we use GPT-4.1 to classify each operation based on its tags (following RQ2 semantic labeling), allowing multiple labels per operation. Operations that cannot be mapped to any category in Table~\ref{tab:rq2_semantic_categories} receive the catch-all label \emph{Other}. We validate this labelling procedure against the manually annotated subset from RQ2 and report multi-label precision, recall, and $F_1$ (precision~$= 0.92$, recall~$= 0.86$, $F_1 = 0.89$). The validated prompt is then applied unchanged to all operations in the RQ3--RQ4 corpus.

\emph{Grouping evaluation.}
After filtering, we apply the Collection/Item Merge pattern (Section~\ref{sec:transformations}) to reduce tool granularity. Because each merged tool replaces two distinct operations with a single interface, we must verify that the new tool preserves the behaviour of both originals. For each of the 232 endpoints in the RQ3 sample that match the Collection/Item Merge pattern, we exercise the merged tool in both modes: once without an identifier argument to test collection retrieval, and once with a valid identifier to test item retrieval. Each invocation is compared against the corresponding original tool output and judged as pass or fail using the same execution criteria as in RQ3. Evaluations were distributed across three authors, and 10\% of merged tools were independently re-assessed by a second author; inter-rater agreement on the pass/fail outcome was perfect (Cohen's~$\kappa = 1.0$).

\vspace{0.5cm}

\section{OpenAPI-to-MCP Generation Framework}
\label{sec:framework}

This section presents our OpenAPI-to-MCP generation framework, which underlies the analyses in RQ3 and RQ4. We distinguish between a baseline generation pipeline and an extended pipeline. The baseline, implemented by  \emph{AutoMCP-Baseline}, is used in RQ3 to assess the extent to which MCP servers can be generated directly from OpenAPI specifications. The extended pipeline, implemented by \emph{AutoMCP}, introduced in RQ4, augments this baseline with specification repair (\textsc{SpecFix}) and tool-set transformations (filtering and groupping) to improve both execution correctness and usability.
\subsection{AutoMCP-Baseline: Compiling OpenAPI into an MCP Server}
\label{sec:automcp}

AutoMCP-Baseline implements the baseline generation pipeline used in RQ3, where each OpenAPI operation is translated into a corresponding MCP tool without additional transformations.

As shown in Algorithm~\ref{alg:automcp}, AutoMCP-Baseline is a static code generator that compiles an OpenAPI~2.0 or~3.0 specification into an executable MCP server stub. Each endpoint in the input specification is transformed into a callable MCP tool, enabling LLM clients to invoke external REST APIs through the Model Context Protocol.

AutoMCP-Baseline is designed for reproducibility and minimal developer intervention: it accepts a specification, performs one-pass offline compilation, and outputs a ready-to-run Python project with environment-based configuration. All operational details, endpoint schemas, parameter structures, and authentication mechanisms,  are inferred directly from the OpenAPI specification. The only required manual step is to provide credentials in the generated \texttt{.env} file.

\begin{algorithm}[h]
\small
\setlength{\lineskip}{0pt}
\setlength{\intextsep}{0pt}
\setlength{\textfloatsep}{4pt}
\setlength{\floatsep}{4pt}
\SetAlgoLined
\SetNlSty{}{}{}
\SetInd{0.2em}{0.8em}
\SetKwInOut{Input}{Require}
\SetKw{KwLet}{let}
\SetKwBlock{Section}{}{}

\caption{\textsc{AutoMCP-Baseline}: OpenAPI $\rightarrow$ MCP server}
\label{alg:automcp}

\Input{OpenAPI spec \texttt{spec}, output dir \texttt{Dout}}

\textbf{A. Parse and load specification} \\
\KwLet{$raw \gets$ \textsc{LoadSpec}$(spec)$} \tcp*[r]{load YAML/JSON}

\textbf{B. Normalize specification} \\
\KwLet{$norm \gets$ \textsc{NormaliseVersion}$(raw)$} \tcp*[r]{unify dialect}
\KwLet{$flat \gets$ \textsc{InlineRefs}$(norm)$} \tcp*[r]{resolve \texttt{\$ref}s}
\textsc{FixMalformedSchemas}$(flat)$ \tcp*[r]{repair common issues}
\KwLet{$eps \gets$ \textsc{ListEndpoints}$(flat)$} \tcp*[r]{extract operations}

\textbf{C. Extract authentication metadata} \\
\KwLet{$sec \gets$ \textsc{ExtractSecurity}$(flat)$} \tcp*[r]{infer auth schemes}
\KwLet{$env \gets$ \textsc{BuildEnvMap}$(sec)$} \tcp*[r]{map credentials to \texttt{.env}}
\If{\textsc{RequiresOAuth2}$(sec)$}{
  \textsc{GenerateOAuth2}$(Dout, sec)$ \tcp*[r]{generate helper if needed}
}

\textbf{D. Generate MCP server stub} \\
\KwLet{$ctx \gets$ \textsc{InitContext}$(Dout, sec, env)$} \tcp*[r]{initialize server context}
\ForEach{$(p, m, op) \in eps$}{
  \KwLet{$par \gets$ \textsc{ExtractParams}$(op)$} \tcp*[r]{extract parameters}
  \KwLet{$par \gets$ \textsc{SanitiseNames}$(par)$} \tcp*[r]{make names safe}
  \KwLet{$h \gets$ \textsc{CreateHandler}$(p, m, par, sec, ctx)$} \tcp*[r]{create tool handler}
  \KwLet{$s \gets$ \textsc{GenerateSchema}$(op)$} \tcp*[r]{derive tool schema}
  \textsc{RegisterTool}$(h, s, ctx)$ \tcp*[r]{register MCP tool}
}

\textbf{E. Emit artifacts} \\
\textsc{WriteEnvFile}$(env, Dout)$ \tcp*[r]{generate \texttt{.env}}
\textsc{SaveGenerated}$(ctx, Dout)$ \tcp*[r]{write server stub}
\end{algorithm}

The compilation process follows the five phases shown in Algorithm~\ref{alg:automcp}.

\underline{\textbf{Step A}}: \textbf{Input parsing and dialect resolution.}
AutoMCP-Baseline parses the input specification and automatically detects the format (YAML or JSON) and dialect (OpenAPI 2.0 or 3.0) by inspecting top-level fields. For OpenAPI 2.0, the base URL is composed from \texttt{schemes}, \texttt{host}, and \texttt{basePath}; for OpenAPI 3.0, the first entry in \texttt{servers} is used. This base URL is injected into all generated tool handlers.

\underline{\textbf{Step B}}: \textbf{Specification normalisation and flattening.}
All \texttt{\$ref} references are recursively resolved to produce a fully inlined version of the specification. AutoMCP-Baseline also performs structural normalisation to reconcile differences between OpenAPI versions, sanitise identifiers, and fix common inconsistencies (e.g., duplicate operation IDs, undefined path parameters).

\underline{\textbf{Step C}}: \textbf{Authentication analysis and \texttt{.env} generation.}
Declared security schemes (e.g., API key, HTTP Basic, Bearer token, OAuth2) are parsed and mapped to environment variables. AutoMCP-Baseline generates a template \texttt{.env} file in the output directory, providing placeholders for all required tokens, secrets, and credentials. If an OAuth2 flow is detected, AutoMCP-Baseline additionally emits a Flask-based \texttt{oauth\_login\_server.py}, which handles the login and callback steps, retrieves access tokens, and writes them into the \texttt{.env} file.

\underline{\textbf{Step D}}: \textbf{Stub generation and handler synthesis.}
Every endpoint is compiled into a Python handler function decorated with \texttt{@mcp.tool(...)}, conforming to the MCP runtime conventions. Handlers validate input, inject authentication headers or query parameters, construct and dispatch HTTP requests using \texttt{requests}, and return the parsed response or raw fallback. When the \texttt{EXTRA\_HEADERS} environment variable is set, AutoMCP-Baseline automatically injects logic to parse a JSON string from the environment and merge it into all HTTP requests, enabling API-specific header customisations.

\underline{\textbf{Step E}}: \textbf{Output layout.}
AutoMCP-Baseline writes all generated artifacts to the specified output directory, including the server stub \texttt{server\_stub.py}, the environment template \texttt{.env}, and optionally \texttt{oauth\_login\_server.py}. The resulting stub runs as an MCP server without additional manual editing.

\underline{\textbf{End-to-End Illustration (Trello API).}}
To illustrate the workflow, we compile Trello's public OpenAPI contract with AutoMCP-Baseline. The generator converts each documented API operation into a corresponding MCP tool and produces handler functions that invoke the underlying REST endpoints. Trello's API-key--based authentication is automatically mapped to environment variables in the generated \texttt{.env} file.

\paragraph{\textbf{Client configuration.}}
Claude Desktop (v2025.6) is configured to launch the generated stub as a local MCP server. Listing~\ref{lst:config} shows the minimal JSON entry required to register the server.

\begin{listing}[h]
\centering
\begin{minipage}{0.75\linewidth}
\caption{Claude configuration for the Trello MCP server generated by AutoMCP-Baseline.}
\label{lst:config}
\begin{lstlisting}[language=json, style=code]
{
  "mcpServers": {
    "trello": {
      "command": "python",
      "args": ["path/to/server_stub.py"]
    }
  }
}
\end{lstlisting}
\end{minipage}
\end{listing}

\begin{figure}[h]
\centering
\includegraphics[width=0.7\linewidth]{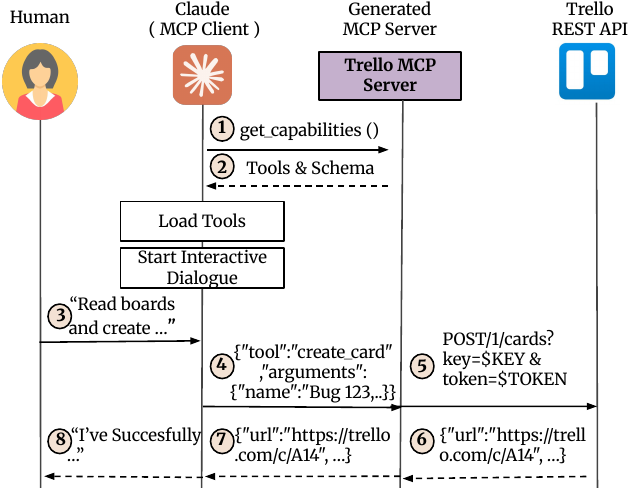}
\caption{End-to-end message flow for Trello card creation mediated by an AutoMCP-Baseline generated MCP server.}
\label{fig:trello_flow}
\end{figure}

\paragraph{\textbf{Runtime flow.}}
At startup, Claude issues \texttt{get\_capabilities}, receives the tool manifest from the generated MCP server, and validates each tool schema. When the user issues a request such as ``Create a bug card in Trello,'' Claude selects the appropriate tool (e.g., \texttt{create\_card}), populates the required JSON parameters, and sends a \texttt{tool\_call}. The server stub injects credentials from the \texttt{.env} file, dispatches the corresponding REST request (e.g., \texttt{POST /1/cards}) to Trello, and returns the structured response. Claude then converts this result into natural-language feedback for the user, as illustrated in Figure~\ref{fig:trello_flow}.

\subsection{SpecFix: Automated OpenAPI Specification Repair}
\label{sec:specfix}

To address the specification-level deficiencies identified in RQ3, we extend the \emph{AutoMCP-baseline} pipeline with \emph{SpecFix}, a preprocessing component for automated OpenAPI contract analysis and repair. SpecFix executes prior to MCP server generation: it compares an OpenAPI specification against its official documentation, detects global inconsistencies, and when enabled, produces a minimally repaired specification for compilation. By separating contract repair from code generation, \emph{AutoMCP-baseline} operates on higher-quality specifications without modifying the underlying compilation logic.

SpecFix targets three categories of specification defects:
\textbf{(A)}~authentication misconfigurations,
\textbf{(B)}~malformed or missing base URLs, and
\textbf{(C)}~undocumented runtime headers or authentication prefixes.
These correspond to global contract inconsistencies that directly prevent
automated invocation of otherwise valid endpoints. The rationale for scoping
SpecFix to these three categories is discussed in
Section~\ref{subsec:rq4-approach}.

SpecFix supports two execution modes. In \emph{analysis mode}, it performs
non-destructive detection and produces a structured issue report. In
\emph{fix mode}, detected issues are validated and, if confirmed, repaired
through targeted structural modifications. The repaired specification preserves
all unaffected content and introduces only the corrections necessary to restore
global contract consistency.

SpecFix takes two inputs: an OpenAPI specification and a URL pointing to the official API documentation. It follows a three-phase architecture, where each phase produces well-defined artifacts consumed by the next.

\underline{\textbf{Phase 1}}: \textbf{Documentation structuring.}
SpecFix constructs a structured representation of the official API
documentation. Because contemporary API portals frequently rely on
client-side rendering frameworks, static HTML retrieval is insufficient.
We therefore employ Playwright in a headless browser environment to execute
JavaScript and capture the fully rendered DOM. During extraction, the crawler
expands collapsible sections, scrolls to trigger lazy-loaded content, and
waits for network-idle conditions so that asynchronous resources are fully
resolved. Extracted pages are normalised and cached to avoid repeated
rendering in subsequent runs. Let $S$ denote the OpenAPI specification and
$D$ the structured documentation model derived from extraction.

\underline{\textbf{Phase 2}}: \textbf{Specification--documentation mismatch detection.}
SpecFix applies a set of deterministic rules that compare structural
properties of the specification $S$ against content extracted from the
documentation model $D$. Each rule targets a specific mismatch pattern
within one of the three supported defect categories (A--C);
Table~\ref{tab:specfix_rules} formalises the six rules used. For
example, an A-type rule fires when the documentation indicates that
authentication is required ($D.\mathit{auth\_mentioned}$) but the
specification declares no security scheme ($\mathit{Sec}_S = \emptyset$).
No LLM is involved at this stage: detection is fully rule-based. For
each triggered rule, SpecFix records the issue type, its location in
the specification, and the supporting specification and documentation
fragments. These records form a structured issue report that serves as
input to Phase~3.

\underline{\textbf{Phase 3}}: \textbf{LLM-based validation and repair.}
In fix mode, SpecFix processes each issue in the report produced by
Phase~2. For each candidate issue, a single call to GPT-4.1 receives
four inputs: the issue type, the relevant specification fragment, the
supporting documentation fragment, and a set of structural constraints
(e.g., valid OpenAPI security scheme formats, required server object
fields). The model first assesses whether the reported inconsistency
is genuine; if rejected, no modification is applied. If confirmed, the
model returns a minimal corrective fragment, the smallest structural
change to the specification that resolves the inconsistency. Each
corrective fragment is associated with a location path in the
specification tree. A deterministic patcher applies the fragment at
the identified location via JSON merge, preserving all unaffected
structure. After all patches are applied, SpecFix serialises the repaired
specification in its original format (YAML or JSON) and records
the applied changes as a structured diff for traceability. 

\begin{table*}[t]
\centering
\caption{SpecFix detection rules for specification--documentation mismatches, grouped by defect category.}
\label{tab:specfix_rules}
\small
\renewcommand{\arraystretch}{0.85}
\begin{tabular}{@{}p{0.17\textwidth}p{0.30\textwidth}p{0.48\textwidth}@{}}
\toprule
\textbf{Issue type} & \textbf{Detection rule} & \textbf{Symbols used} \\
\midrule
\multicolumn{3}{@{}l}{\textit{(A) Authentication misconfiguration}} \\
\midrule
Missing security declaration &
$D.\mathit{auth\_mentioned} \wedge \mathit{Sec}_S = \emptyset$ &
$\boldsymbol{\mathit{Sec}_S}$: declared security schemes in specification $S$; \newline
$\boldsymbol{D.\mathit{auth\_mentioned}}$: documentation indicates that authentication is required. \\
Inconsistent security scheme type &
$\mathit{Sec}_S \neq \emptyset \wedge
\exists \sigma \in \mathit{Sec}_S :
\neg \mathit{mentioned}_D(\sigma.\mathit{type})$ &
$\boldsymbol{\sigma}$: a declared security scheme; \newline
$\boldsymbol{\mathit{mentioned}_D(x)}$: documentation explicitly mentions authentication type $x$. \\
\midrule
\multicolumn{3}{@{}l}{\textit{(B) Malformed or missing base URL}} \\
\midrule
Missing server configuration &
$(\mathit{Swagger2}(S) \wedge S.\mathit{host} = \emptyset)
\vee
(\mathit{OpenAPI3}(S) \wedge \mathit{Serv}_S = \emptyset)$ &
$\boldsymbol{\mathit{Swagger2}(S)}$: $S$ conforms to Swagger 2.x; \newline
$\boldsymbol{\mathit{OpenAPI3}(S)}$: $S$ conforms to OpenAPI 3.x; \newline
$\boldsymbol{\mathit{Serv}_S}$: declared server configuration in $S$. \\
Malformed or invalid base URL &
$\exists srv \in \mathit{Serv}_S :
\neg \mathit{valid\_url}(srv.\mathit{url})$ &
$\boldsymbol{srv}$: a declared server entry; \newline
$\boldsymbol{\mathit{valid\_url}(u)}$: $u$ is a valid base URL (protocol present, no unresolved template variables). \\
\midrule
\multicolumn{3}{@{}l}{\textit{(C) Undocumented runtime headers or authentication prefixes}} \\
\midrule
Missing required global header &
$\exists h \in D.\mathit{global\_headers} : h \notin S$ &
$\boldsymbol{D.\mathit{global\_headers}}$: headers documented as globally required; \newline
$\boldsymbol{h}$: a documented required header. \\
Authentication prefix mismatch &
$D.\mathit{auth\_prefix}\ \text{defined} \wedge
\mathit{prefix}_S \neq D.\mathit{auth\_prefix}$ &
$\boldsymbol{D.\mathit{auth\_prefix}}$: documented authentication prefix (e.g., \texttt{Bearer}); \newline
$\boldsymbol{\mathit{prefix}_S}$: authentication prefix declared in the specification. \\
\bottomrule
\end{tabular}
\end{table*}

\subsection{Tool-Set Transformations: Filtering and Grouping}
\label{sec:transformations}

AutoMCP-Baseline translates each OpenAPI operation into a separate MCP
tool. For large APIs, this one-to-one mapping can produce hundreds of
tools, exceeding the context-window and tool-selection capacity of
current LLMs. To address this, AutoMCP incorporates two
transformations---category-based filtering and deterministic
grouping---grounded in the empirical patterns identified in RQ2.

\underline{\textbf{Step 1}}: \textbf{Category-based filtering.}
Each OpenAPI operation is labelled with the structural and semantic
categories defined in RQ2 (Tables~\ref{tab:rq2_struct_results}
and~\ref{tab:rq2_semantic_categories}). Semantic labels are generated
by GPT-4.1 from operation tags, allowing an operation to receive
multiple labels when applicable. Operations that cannot be mapped to
any known category are assigned the label ``Other.'' To determine
which categories to filter, we derive a data-driven threshold from
the omission statistics in RQ2. Rather than using a fixed cutoff, we
compute the 75th percentile ($Q_{75}$) of the omission distribution
across all feature categories. A category $c$ is considered
\emph{highly omitted} if either its macro omission rate or its micro
omission rate exceeds the respective $Q_{75}$ threshold:
\begin{equation}
\mathit{MO}(c) \ge Q_{75}^{\mathit{MO}} \;\lor\; \mathit{mO}(c) \ge Q_{75}^{\mathit{mO}}
\end{equation}
This rule yields six filtered categories: four semantic---authorization,
authentication, settings/configuration, and user management---and two
structural---deprecated and destructive operations. An operation is
removed only if \emph{all} of its assigned labels fall within this
union, preserving operations that serve multiple purposes including
at least one commonly exposed capability.

\underline{\textbf{Step 2}}: \textbf{Collection/Item Merge grouping.}
After filtering, we reduce tool granularity by applying the
Collection/Item Merge pattern (Table~\ref{tab:aggregation_patterns}),
the most frequent many-to-one aggregation pattern observed in RQ2.
For each API, we scan the remaining operations and identify pairs of
retrieval endpoints of the form \texttt{GET /r} (collection) and
\texttt{GET /r/\{id\}} (item), where \texttt{r} denotes the same
normalised resource path prefix and \texttt{\{id\}} is a single
path-parameter segment. For each detected pair, we synthesise a
single MCP tool with an optional identifier argument: when the
identifier is omitted, the tool executes the collection endpoint;
when provided, it executes the item endpoint.\

\subsection{AutoMCP: End-to-End Pipeline}
\label{sec:pipeline}

\begin{figure}[h]
\centering
\includegraphics[width=0.95\linewidth]{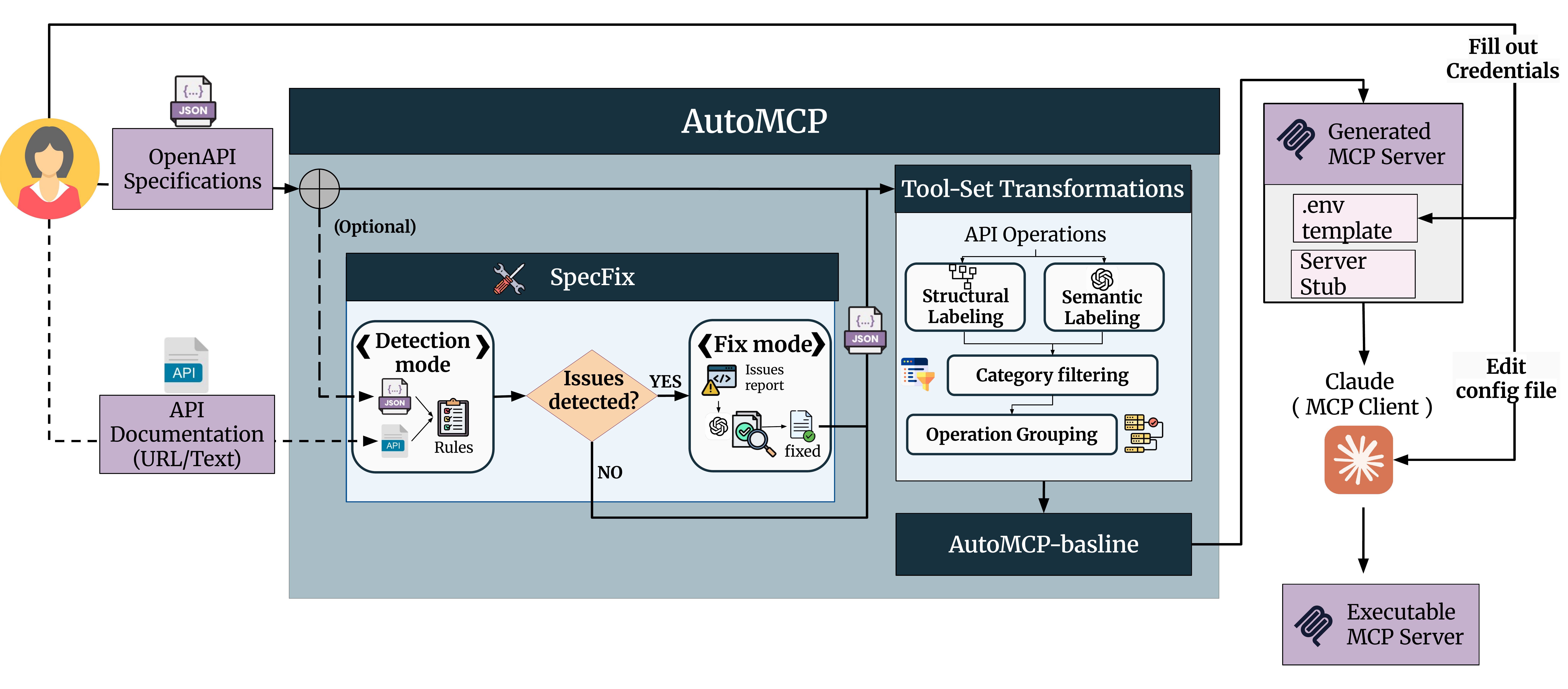}
\caption{AutoMCP end-to-end pipeline. Optional stages are indicated with dashed borders.}
\label{fig:pipeline}
\end{figure}

AutoMCP integrates the three components described above---baseline
generation (Section~\ref{sec:automcp}), specification repair
(Section~\ref{sec:specfix}), and tool-set transformations
(Section~\ref{sec:transformations}), into a single pipeline that
compiles an OpenAPI specification into an executable MCP server.
Figure~\ref{fig:pipeline} illustrates the end-to-end workflow.

The pipeline takes two inputs: an OpenAPI specification describing a
vendor API and, optionally, a URL pointing to its official
documentation. When a documentation URL is provided, SpecFix analyses
the specification for contract-level mismatches and, in fix mode,
produces a minimally repaired specification
(Section~\ref{sec:specfix}). The repaired (or original) specification
then enters the transformation stage, where category-based filtering
removes operations belonging to highly omitted categories and
Collection/Item Merge grouping consolidates endpoint
pairs (Section~\ref{sec:transformations}). The resulting operation set
is compiled into MCP tools using the AutoMCP-Baseline code generation
procedure (Section~\ref{sec:automcp}). The generated server exposes
these tools via the MCP protocol and can be invoked by any
MCP-compatible client.

AutoMCP operates in three configurations of increasing capability,
each corresponding to a specific evaluation stage in this study:

\begin{itemize}
\item \textbf{AutoMCP-Baseline} (RQ3): direct one-operation-to-one-tool
compilation, with no repair or transformations.
\item \textbf{AutoMCP-baseline + SpecFix} (RQ4, Stage~1): specification repair
followed by baseline compilation.
\item \textbf{AutoMCP-baseline + SpecFix + Transformations} (RQ4, Stage~2):
specification repair, category-based filtering, Collection/Item Merge
grouping, and compilation.
\end{itemize}

%% file: Sections/ResultsAndAnalysis.tex

\section{Results }
\label{sec:evaluation}

This section presents the empirical findings for all research questions (RQ1–RQ4).

\subsection{RQ1: REST reliance, integration, and configuration in MCP servers}
\label{sec:rq1-results}

\subsubsection{RQ1.1 REST Reliance}

Across the 116 official MCP servers in our corpus, 114 expose tool configurations that can be statically enumerated and analyzed. We excluded two servers\footnote{\texttt{integromat/make-mcp-server} and \texttt{PipedreamHQ/pipedream}} that generate tool sets dynamically from user-specific configurations, preventing reproducible tool enumeration and classification.

\textbf{Most official MCP servers realize tool functionality through vendor REST API calls.} Using the Stage~1 procedure (Section~\ref{sec:rq1-approach}), we find that 87 servers (76.3\%) are REST-backed, meaning that all tool handlers can be statically traced to one or more vendor REST API invocations. These invocations occur via direct HTTP calls, SDKs wrapping REST endpoints, or equivalent REST clients. An additional 14 servers (12.3\%) are partially REST-backed, where only a subset of tools relies on REST invocations. The remaining 13 servers (11.4\%) are \emph{non-REST-backed}, with tools relying on alternative mechanisms such as direct database connections (e.g., SQL drivers, gRPC), GraphQL APIs, binary protocols (e.g., Bolt, Lightning/NWC), or local application integrations (e.g., WebSocket, browser automation). 

\begin{table}[h]
\centering
\small
\setlength{\tabcolsep}{6pt}
\begin{tabularx}{0.8\linewidth}{llll}
\toprule
\textbf{MCP Server Class} & \textbf{Servers} & \textbf{Total Tools} & \textbf{Median Tools/server} \\
\midrule
REST-backed & 87 (76.3\%) & 2{,}636 (83.3\%) & 18 \\
Partially REST-backed & 14 (12.3\%) & 347 (11\%) & 21 \\
Non-REST-backed & 13 (11.4\%) & 180 (5.7\%) & 11 \\
\midrule
\textbf{Total} & 114 (100\%) & 3,163 (100\%) & 18 \\
\bottomrule
\end{tabularx}
\caption{Corpus distribution of official MCP servers ($n=114$).}
\label{tab:rq1_corpus}
\end{table}

\begin{figure}[tbp]
  \centering
  \includegraphics[width=\linewidth]{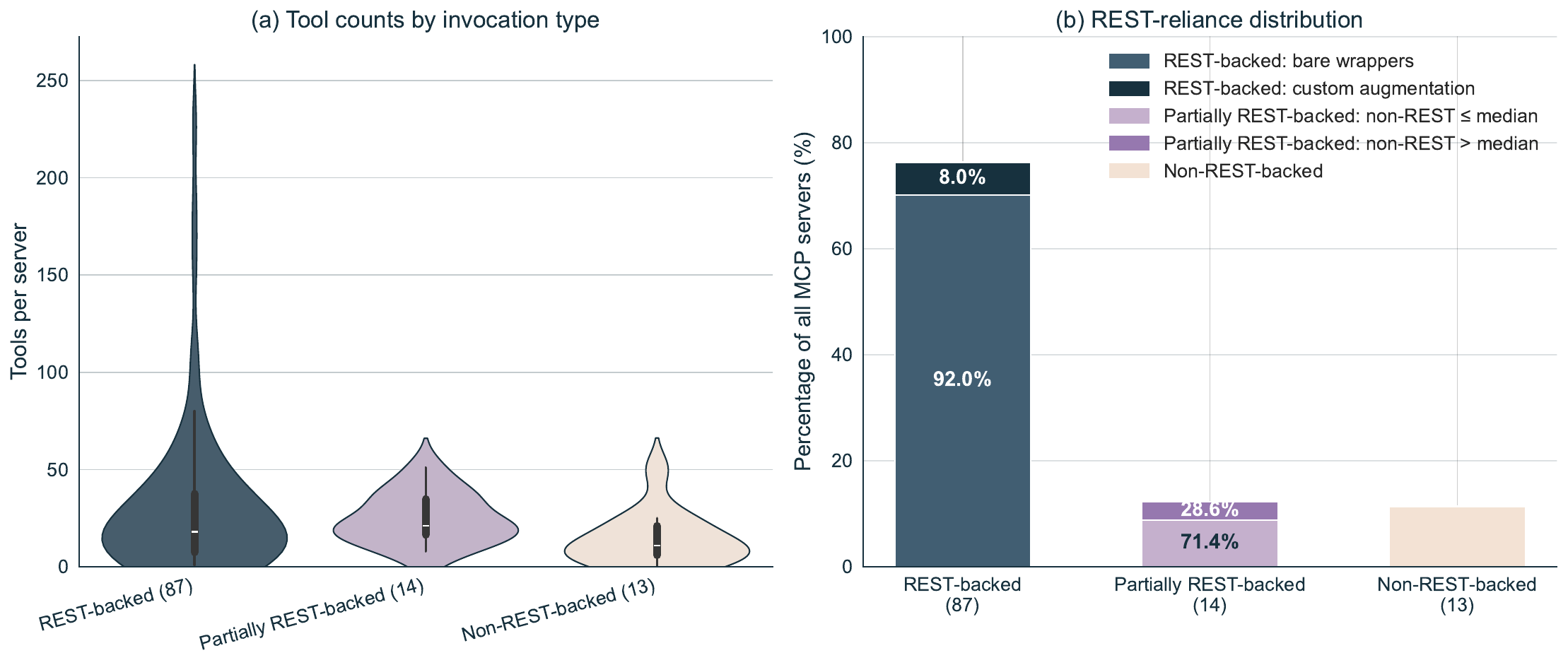}
  \caption{Distribution of MCP servers by REST-reliance and tool counts.
  (\textbf{a}) Tool count distribution (violin plot).
  (\textbf{b}) REST-reliance breakdown (stacked bars).}
  \label{fig:combined}
\end{figure}

Table~\ref{tab:rq1_corpus} and \autoref{fig:combined} summarizes the distribution at both repository and tool levels. REST-backed servers dominate the ecosystem, while partially and non-REST-backed servers are comparatively less common. Tool counts vary substantially across classes: REST-backed servers exhibit the widest distribution, reflecting variation in the size of exposed vendor APIs, whereas partially and non-REST-backed servers tend to expose smaller and more uniform tool sets.

\textbf{Departures from REST are comparatively rare and concentrated in a small set of recurring mechanisms.} In partially REST-backed servers, non-REST functionality is typically confined to a minority of tools (median share 22.5\%, range 4.3\%–90.9\%). Only four repositories implement more than one quarter of their tools via non-REST mechanisms (PortSwigger~\cite{portswigger_mcp}, Neo4j~\cite{neo4j_mcp}, MongoDB~\cite{mongodb_mcp}, and Neon~\cite{neon_mcp}). Across the 14 partially REST-backed servers, we identified 110 non-REST tools in total. These tools fall into a small set of recurring non-REST mechanism families. The largest group uses database-native interfaces (49.1\%), where tools interact through database drivers or query/protocol stacks rather than vendor HTTP endpoints. A second group relies on local or in-process interfaces (18.2\%), in which tools invoke functionality via locally running applications, plugins, or embedded libraries instead of remote REST calls. A third group uses RPC-style service interfaces (10.0\%). The remaining tools rely on a small number of vendor-specific mechanisms, without forming consistent or widely adopted patterns.

\begin{findingbox}
\textbf{\faKey\ Finding-RQ1.1.} Vendor REST APIs overwhelmingly dominate MCP server implementations: 88.6\% of servers in our corpus ($n=114$) are fully or partially REST-backed. Non-REST mechanisms are rare and concentrated in a small number of specialized integration patterns.
\end{findingbox}

\subsubsection{RQ1.2 Configuration and Integration Practices}

\paragraph{(a) Configuration practices.}
We now analyze how REST-backed servers configure authentication, communication, and tool exposure at runtime. Table~\ref{tab:mcp-auth} summarizes authentication mechanisms, while Tables~\ref{tab:mcp-transport} and~\ref{tab:mcp-selection} report transport support and tool-selection capabilities.

\textbf{Authentication is dominated by token-based credential provisioning.}
API-key authentication dominates (48.3\%), followed by HTTP bearer tokens (17.2\%) and OAuth~2.0 (10.3\%), as shown in Table~\ref{tab:mcp-auth}. A smaller share of servers rely on multi-credential configurations (12.6\%), such as API key and secret pairs or client ID/secret combinations. The remaining servers (10.3\%) use alternative approaches, including cloud credential chains or CLI-based authentication (e.g., AWS, Azure, and Google Cloud identity systems). Overall, MCP servers predominantly rely on non-interactive, token-based authentication that mirrors the credential models of the underlying vendor APIs.

\begin{table}[h]
\centering
\small
\caption{Authentication mechanisms used by MCP servers ($n=87$).}
\label{tab:mcp-auth}
\begin{tabular}{lrr}
\toprule
\textbf{Authentication type} & \textbf{Servers} & \textbf{\%} \\
\midrule
API key & 42 & 48.3\% \\
HTTP Bearer & 15 & 17.2\% \\
OAuth 2.0 & 9 & 10.3\% \\
HTTP Basic & 1 & 1.1\% \\
Multi-credential & 11 & 12.6\% \\
Other (cloud IAM / CLI auth) & 9 & 10.3\% \\
\midrule
\textbf{Total} & \textbf{87} & \textbf{100\%} \\
\bottomrule
\end{tabular}
\end{table}

\textbf{STDIO remains the modal transport, with substantial multi-transport support.}
STDIO-only transport is the most common (50.6\%), reflecting the prevalence of local, subprocess-style deployment in MCP integrations. In contrast, 40.2\% of servers support multiple transports, typically combining STDIO with HTTP and/or SSE endpoints. HTTP-only deployments account for 8.0\%, while SSE-only configurations are rare (1.1\%). This distribution suggests a split between servers optimized for local execution and those designed for flexible deployment across runtime environments.

\begin{table}[h]
\centering
\small
\caption{Transport mechanisms supported by MCP servers ($n=87$).}
\label{tab:mcp-transport}
\begin{tabular}{lrr}
\toprule
\textbf{Transport type} & \textbf{Servers} & \textbf{\%} \\
\midrule
STDIO only & 44 & 50.6\% \\
Multiple transports & 35 & 40.2\% \\
HTTP only & 7 & 8.0\% \\
SSE only & 1 & 1.1\% \\
\midrule
\textbf{Total} & \textbf{87} & \textbf{100\%} \\
\bottomrule
\end{tabular}
\end{table}

\textbf{Explicit tool filtering is uncommon, and most servers expose all tools by default.}
Only 26.4\% of MCP servers provide configuration mechanisms to enable, disable, or selectively register subsets of tools (e.g., via flags such as \texttt{--tools}, \texttt{--enabled-tools}, or environment variables like \texttt{MCP\_ENABLED\_TOOLS}), as shown in Table~\ref{tab:mcp-selection}. The majority (73.6\%) expose all tools unconditionally at initialization. This suggests that current implementations prioritize simplicity and immediate tool availability over configurable capability scoping, although the presence of filtering in a non-trivial minority indicates emerging attention to manageability.

\begin{table}[t]
\centering
\small
\caption{Tool-selection capability among MCP servers ($n=87$).}
\label{tab:mcp-selection}
\begin{tabular}{lrrr}
\toprule
\textbf{Tool selection support} & \textbf{Servers} & \textbf{\%} & \textbf{Median tools/server} \\
\midrule
Yes (tools can be filtered) & 23 & 26.4\% & 35 \\
No (all tools exposed) & 64 & 73.6\% & 16 \\
\midrule
\textbf{Total} & \textbf{87} & \textbf{100\%} & -- \\
\bottomrule
\end{tabular}
\end{table}

\begin{findingbox}
\textbf{\faKey\ Finding-RQ1.2.a.}
REST-backed MCP servers prioritize low-friction configuration: token-based authentication dominates, STDIO is the most common transport, and most servers expose all tools by default.
\end{findingbox}

\paragraph{(b) Integration strategies.}
We now examine how these servers incorporate REST calls into tool implementations.

\textbf{\emph{Bare API consumption} is the dominant integration strategy in REST-backed MCP servers.} Among the 87 REST-backed servers, 80 repositories (92.0\%) consist exclusively of tools that perform bare API consumption. Across all REST-backed MCP servers, we identified 2,636 tools, of which 2,436 (92.4\%) belong to bare-consumption servers. 
\begin{table}[h]
\centering
\small
\begin{tabular}{p{4cm} r r}
\toprule
\textbf{Invocation type} & \textbf{Tools} & \textbf{\%} \\
\midrule
Direct HTTP & 1,579 &  64.8\% \\
SDK-based  & 705 &  28.9\% \\
OpenAPI-generated/client & 111 &  4.6\% \\
CLI-based & 41 &  1.7\% \\
\midrule
\textbf{Total} & 2,436 &  100\% \\
\bottomrule
\end{tabular}
\caption{Invocation type distribution among \emph{Bare API consumption} MCP servers and tools.}
\label{tab:rest-invocation}
\end{table}

\textbf{Direct HTTP requests dominate how \emph{bare API consumption} tools call vendor APIs.} Among the 2,436 bare-consumption tools, direct HTTP invocation is the most prevalent mechanism (1,579 tools; 64.8\%), as shown in Table~\ref{tab:rest-invocation}. SDK-based invocation is the second most frequent mechanism (705 tools; 28.9\%), while OpenAPI-generated clients (111 tools; 4.6\%) and CLI-based invocation (41 tools; 1.7\%) are comparatively rare.

\textbf{Custom API client logic in REST-backed MCP servers is rare and concentrated in a small minority of REST-backed servers.}
Only 7 of the 87 REST-backed repositories (8.0\%) introduce non-trivial augmentation logic beyond \emph{bare API consumption} and are classified as \emph{custom API client}. These 7 repositories account for 200 tools (7.7\% of REST-backed tools). In all cases, tools still ultimately invoke vendor REST endpoints; the distinguishing characteristic is the presence of additional computation or coordination layered on top of routine API access.
\begin{table}[h]
\centering
\caption{ Taxonomy of augmentation patterns in \emph{Custom API client} servers (7 of 87 REST-backed servers). Patterns are non-mutually exclusive.}
\label{tab:augmentation-taxonomy-min}
\footnotesize
\begin{tabular}{p{3.2cm}p{7.0cm}rr}
\toprule
\textbf{Pattern} & \textbf{Definition (beyond routine request processing)} & \textbf{\#Servers} & \textbf{\#Tools} \\
\midrule
Workflow orchestration &
A tool coordinates multiple upstream interactions to produce one tool result (e.g., multi-call composition, aggregation, and asynchronous completion via polling). &
4 & 63 \\

AI-mediated transformation &
A tool invokes model-driven processing to synthesize or restructure artifacts (e.g., schema/artifact generation, assistant-style translation, or retrieval-augmented suggestion). &
5 & 28 \\

Non-functional controls &
A tool introduces execution semantics affecting performance/safety (e.g., caching, rate limiting, response sanitization or prompt injection mitigation). &
3 & 83 \\
\bottomrule
\end{tabular}
\end{table}

\textbf{Augmentation behaviors in Custom API client tools cluster into a small set of recurring patterns.}
Static inspection of \emph{Custom API client} tools reveals three recurring augmentation patterns that go beyond routine client REST API responsibilities (e.g., authentication handling, request formatting, pagination traversal, and error processing). Table~\ref{tab:augmentation-taxonomy-min} summarizes the taxonomy and its prevalence across servers. \emph{Workflow orchestration} composes multiple upstream calls into a single tool-level result, handling inter-call dependencies such as polling and aggregation internally. \emph{AI-mediated transformation} interposes a secondary model invocation within the tool. \emph{Non-functional controls} embed cross-cutting safeguards (caching, rate limiting, response sanitization) directly in the tool logic. The three patterns differ markedly in adoption granularity: non-functional controls and workflow orchestration, once adopted, are applied broadly across a server's tool surface (83 and 63 tools, respectively), whereas AI-mediated transformation is deployed selectively at specific endpoints (28 tools across 5 servers).

\begin{lstlisting}[
language=JavaScript,
backgroundcolor=\color{codebg},
xleftmargin=0.5cm,
xrightmargin=0.5cm,
basicstyle=\ttfamily\footnotesize,
breaklines=true,
breakatwhitespace=true,
columns=fullflexible,
keepspaces=true,
keywordstyle=\color{blue}\bfseries,
stringstyle=\color{codepurple},
commentstyle=\color{codegreen}\itshape,
escapeinside={(*@}{@*)},
identifierstyle=\color{black},
emph={addTool, url, extraction_prompt, scrape_response, markdown_content, sampling_response, system_prompt, user_prompt},
emphstyle=\color{codered},
emph={[2]const, let, async, await, return, name, description},
emphstyle={[2]\color{blue}\bfseries},
caption={BrightData MCP example \cite{brightdata_mcp} with direct HTTP integration and AI-mediated transformation augmentation.},
label={lst:ai-augmentation}
]
// MCP TOOL: extract
addTool({
  name: 'extract',
  description: 'Scrape webpage and extract structured JSON.',
  parameters: z.object({
    url: z.string().url(),
    extraction_prompt: z.string().optional(),
  }),
  execute: tool_fn('extract', async ({ url, extraction_prompt }, ctx) => {

    let scrape_response = await axios({                       // API call
      url: 'https://api.brightdata.com/request',
      method: 'POST',
      data: { url, format: 'markdown' },
    });

    let markdown_content = scrape_response.data;

    let system_prompt = "Return ONLY valid JSON.";
    let user_prompt = extraction_prompt || "Extract structured data.";

    let sampling_response = await session.requestSampling({  // AI augmentation
      messages: [{
        role: "user",
        content: { type: "text",
          text: `${user_prompt}\n\n${markdown_content}` },
      }],
      systemPrompt: system_prompt,
    });

    return sampling_response.content.text;
  }),
});
\end{lstlisting}

Listing~\ref{lst:ai-augmentation} illustrates one tool from the BrightData MCP server, which uses \emph{direct HTTP invocation} to retrieve webpage content in markdown format. However, the tool does not simply return the API response. Instead, it introduces additional logic by invoking an LLM through a sampling interface, using system and user prompts to transform the retrieved content into structured JSON. As captured in Table~\ref{tab:augmentation-taxonomy-min}, this corresponds to \emph{AI-mediated transformation}.

\textbf{Augmentation intensity in Custom API client servers is highly skewed, indicating both repository-wide and targeted adoption modes.}
As illustrated in \autoref{tab:augmentation-per-api}, within the 7 \emph{custom API client} repositories, augmentation is not uniformly applied: 5 server repositories augment more than 95\% of their tools, whereas the remaining 2 servers apply augmentation to only a small fraction (less than 5\%) of an otherwise bare tool set. This skew suggests that augmentation is typically adopted either as a repository-wide design stance (systematically layering orchestration or controls across tools) or as a targeted extension used for a small number of specialized tools.

\begin{table}[htbp]
\centering
\caption{Augmentation in \emph{Custom API client} MCP servers (7 of 87 REST-backed servers). }
\label{tab:augmentation-per-api}
\footnotesize
\begin{tabular}{lrrr@{\hspace{1em}}p{6.2cm}}
\toprule
\textbf{MCP Server} & \textbf{\#Tools} & \textbf{\#Aug. tools} & \textbf{\% Aug.} & \textbf{Augmentation patterns} \\
\midrule
Apify \cite{apify_mcp} & 17 & 17 & 100 & Workflow orchestration; AI-mediated transformation \\
Honeycomb \cite{honeycomb_mcp} & 14 & 14 & 100 & Non-functional controls; Workflow orchestration \\
Astra DB \cite{astra_mcp} & 16 & 16 & 100 & Non-functional controls \\
AWS MCP \cite{openapi_mcp_server} & 2 & 2 & 100 & Workflow orchestration; AI-mediated transformation \\
BrightData \cite{brightdata_mcp} & 63 & 60 & 95 & Workflow orchestration; AI-mediated transformation;Non-functional controls \\
Cashfree \cite{cashfree_mcp} & 39 & 2 & 5.1 & AI-mediated transformation \\
Firebase \cite{firebase_mcp} & 49 & 2 & 4.1 & AI-mediated transformation \\
\midrule
\textbf{Total} & 200 & 113  & --- & \\
\bottomrule
\end{tabular}
\end{table}

\begin{findingbox}
\textbf{\faKey\ Finding-RQ1.2.b.}
REST-backed MCP servers act as thin adapters over vendor APIs: 92\% implement tools as lightweight REST pass-throughs, with functionality realized through bare API consumption, predominantly via direct HTTP. Augmentation beyond routine API access is rare (8\%) and, when present, follows a small set of recurring patterns: workflow orchestration, AI-mediated transformation, and non-functional controls.
\end{findingbox}

\subsection{RQ2: REST operation exposure, omission, and mapping patterns in MCP tools}

\subsubsection{{RQ2.1: How much of the vendor REST API surface do MCP servers expose?}}\mbox{}\

Across the 42 MCP servers paired with publicly available OpenAPI specifications, we extracted a total of 6,966 REST operations. Using the alignment procedure described in   section \ref{subsec:rq2-approach}, we mapped 1,013 REST operations to 968 MCP tools. 

\begin{figure}[h]
    \centering
    \begin{subfigure}[b]{0.45\textwidth}
        \centering
        \includegraphics[width=\textwidth]{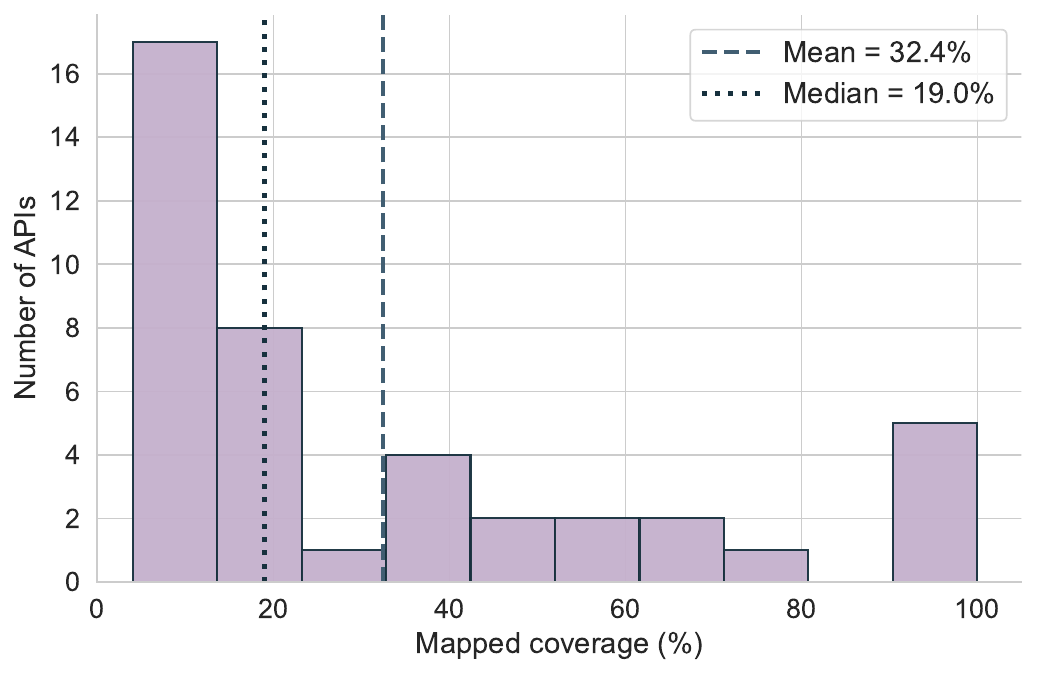}
        \caption{Distribution of per-server coverage ratios.}
    \end{subfigure}
    \hfill
    \begin{subfigure}[b]{0.45\textwidth}
        \centering
        \includegraphics[width=\textwidth]{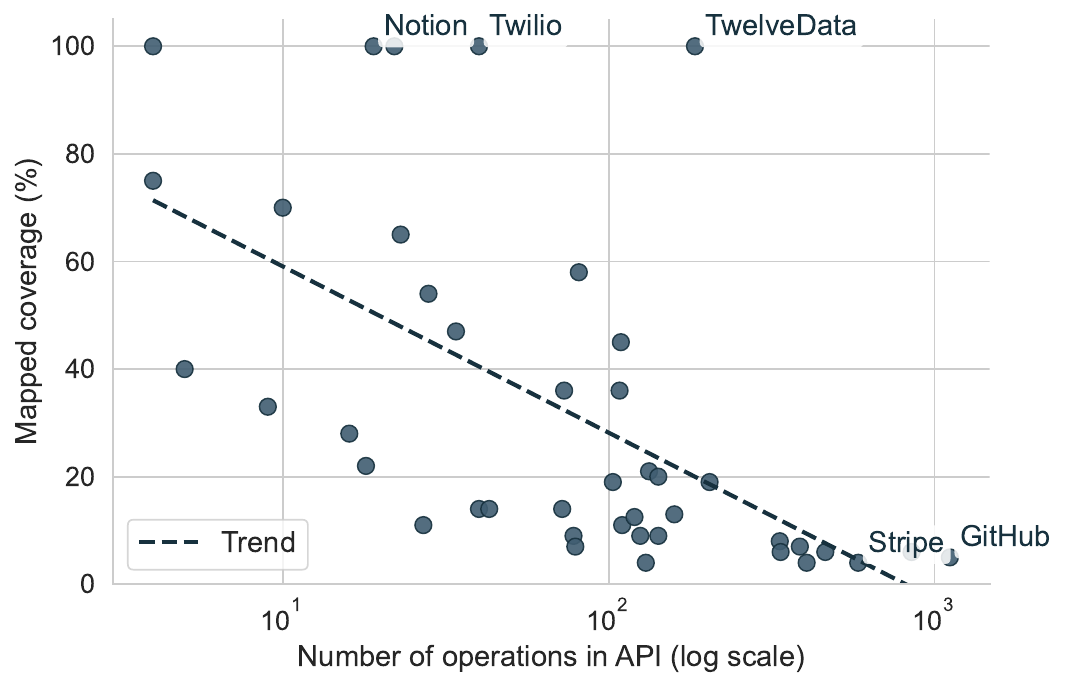}
        \caption{Coverage ratio as a function of vendor API size (log scale).}
    \end{subfigure}
    \hfill
\caption{Distribution and scaling of REST operation coverage across matched MCP servers.}    
\label{fig:rq2-coverage}
\end{figure}

\textbf{Most MCP servers expose fewer than one in five operations available in vendor OpenAPI specifications.}
Across APIs, the median operation-level coverage is 19\%, indicating that MCP servers typically surface only a subset of the vendor interface. The mean coverage is higher (32.4\%), indicating a right-skewed distribution(\autoref{fig:rq2-coverage}a): a few servers expose a large share of their vendor's operations, while the majority surface only a narrow subset.

\textbf{MCP servers expose proportionally fewer operations as vendor API size increases.}\autoref{fig:rq2-coverage}b plots coverage against API size, showing a clear inverse trend. We partition APIs into size categories using box-plot quartiles~\cite{tukey1977eda}. We partition APIs into size categories using box-plot quartiles~\cite{tukey1977eda}: \emph{small} ($\leq Q_{1}$, i.e., $\leq 25$ operations), \emph{medium} ($Q_{1}$--$Q_{2}$), \emph{large} ($Q_{2}$--$Q_{3}$), and \emph{very large} ($> Q_{3}$, i.e., $> 300$ operations). For small APIs ($n=10$), the median coverage is 68\% (mean $= 63\%$). Coverage declines sharply with scale: medium APIs ($n=11$) exhibit a median of 14\% (mean $= 33\%$), large APIs ($n=13$) a median of 19\% (mean $= 18\%$), and very large APIs ($n=8$) drop to a median of only 6\%. Only 5 APIs (12\%) expose every REST operation through MCP tools.

\begin{findingbox}
\textbf{\faKey\ Finding-RQ2.1.} MCP servers act as curated views over vendor APIs rather than complete proxies, with coverage driven primarily by API size: larger API surfaces yield proportionally narrower exposure.
\end{findingbox}

\subsubsection{{RQ2.2: Which categories of REST operations are preferentially exposed or omitted?}}\mbox{}\
\paragraph{(a) Structural Feature Analysis}
We now report coverage of operations based on structural features. Table~\ref{tab:rq2_struct_results} reports macro/micro-average coverage and lift across the 9 structural categories.

\begin{table*}[h]
\centering
\small
\resizebox{\textwidth}{!}{
\begin{tabular}{lrrrrrrr}
\toprule
\textbf{Structural feature} &
\textbf{Macro Cov} &
\textbf{Macro Omit} &
\textbf{Macro Lift} &
\textbf{\# APIs} &
\textbf{Micro Cov} &
\textbf{Micro Omit} &
\textbf{\# Ops} \\
\midrule
is\_read\_only        
& 0.3816 & 0.6184 & \cellcolor{codeem!35}1.2596 & 42 & 0.1940 & 0.8060 & 3418 \\

is\_listing\_operation
& 0.4122 & 0.5878 & \cellcolor{codeem!35}1.4470 & 40 & 0.1847 & 0.8153 & 1538 \\

is\_mutating          
& 0.2805 & 0.7195 & 0.8694 & 39 & 0.1094 & 0.8906 & 2679 \\

is\_destructive       
& 0.2041 & \cellcolor{codebg!90!orange}0.7959 & 0.6347 & 32 & 0.0657 & \cellcolor{codebg!90!orange}\textbf{0.9343} & 867 \\

needs\_authentication 
& 0.3555 & 0.6445 & 0.9989 & 41 & 0.1543 & 0.8457 & 5949 \\

deep\_path            
& 0.2785 & 0.7215 & 0.7983 & 38 & 0.1115 & 0.8885 & 2851 \\

has\_request\_body    
& 0.2964 & 0.7036 & 0.9277 & 37 & 0.1009 & 0.8991 & 2536 \\

has\_pagination       
& 0.2995 & 0.7005 & 1.0763 & 23 & 0.2544 & 0.7456 & 456 \\

is\_deprecated        
& 0.1074 & \cellcolor{codebg!90!orange}\textbf{0.8926} & 0.6592 & 13 & 0.0531 & \cellcolor{codebg!90!orange}\textbf{0.9469} & 226 \\

\midrule
\textbf{Median} 
& \textbf{0.2964} & \textbf{0.7036} & \textbf{0.9277} & \textbf{38} 
& \textbf{0.1115} & \textbf{0.8885} & \textbf{2536} \\

\textbf{Q75}    
& \textbf{0.3555} & \textbf{0.7215} & \textbf{1.0763} & \textbf{40} 
& \textbf{0.1847} & \textbf{0.8991} & \textbf{2851} \\

\bottomrule
\end{tabular}
}
\caption{Structural feature coverage, omission, and lift across APIs. Cells highlighted in \textcolor{codeem}{blue} indicate high values (above the 75th percentile, Q75), while cells highlighted in orange indicate the highest omission values.}
\label{tab:rq2_struct_results}
\end{table*}

\textbf{MCP servers preferentially expose retrieval operations and omit state-changing and deprecated functionality.}
As illustrated in \autoref{tab:rq2_struct_results}, read-only and listing operations are the only features with macro lift above 1 (1.26 and 1.45, respectively), confirming that retrieval operations are more likely to be included than other operation types. Mutating and destructive operations fall below baseline (lift of 0.87 and 0.63), with destructive operations exhibiting the highest micro omission rate among non-deprecated features (0.93). Deprecated operations are near-universally excluded (macro omission 0.89, micro omission 0.95), suggesting that deprecation markers in OpenAPI specifications serve as effective exclusion signals. 


\textbf{Structurally complex operations tend to be under-exposed, with pagination as a notable exception.}
Operations that signal invocation complexity namely, deep paths and request bodies, are associated with below-baseline lift (0.80 and 0.93, respectively), suggesting that MCP server developers favour simpler endpoint signatures. Pagination-bearing operations are the sole exception: despite appearing in only 23 APIs, they exhibit above-baseline lift (1.08) and the highest micro coverage among all features (0.25), likely because paginated endpoints correspond to high-value listing operations that developers prioritise regardless of their added complexity.

\begin{findingbox}
\textbf{\faKey\ Finding-RQ2.2.1.}  
MCP servers are more likely to expose operations related to information retrieval, while selectively omitting operations that are either higher impact (mutating/destructive), structurally complex (deep paths, request bodies), or explicitly marked as deprecated.
\end{findingbox}

\begin{table*}[h]
\centering
\small
\setlength{\tabcolsep}{3.5pt}
\renewcommand{\arraystretch}{0.8}

\begin{tabularx}{\textwidth}{
>{\raggedright\arraybackslash}p{0.2\textwidth}
>{\raggedright\arraybackslash}X
>{\centering\arraybackslash}p{0.08\textwidth}
}
\toprule
\textbf{Semantic category} & \textbf{Representative tags} & \textbf{\# APIs} \\
\midrule
Inventory management &
billing, quota, invoices, products, pricing, payments, exchanges, inventory, assets, vendors &
20 \\

Data / storage &
search, database, query, migrations, datasets, indexes, export, statements, federation &
17 \\

Authentication &
auth, oauth, tokens, credentials, access tokens, API keys, federated authentication, grants &
17 \\

Event \& workflow management &
events, schedules, triggers, actions, workflows, automation, health events, orchestration &
17 \\

User management &
users, accounts, service accounts, admin users, account members, user tags, bundles &
17 \\

License / policy / requirements &
licensing, regulatory requirements, approvals, compliance, policies, standards, disputes &
16 \\

Analytics &
metrics, analytics, time series, correlations, analysis, insights, trending, statistics &
15 \\

Authorization &
roles, permissions, access control, RBAC, allowlist, domains, role bindings &
15 \\

Settings / configuration &
settings, SSO, preferences, configuration, session settings, sandbox, tuning &
11 \\

Key management &
encryption keys, signing keys, key management, API keys, keystore, BYOK, certificates &
11 \\

Logs &
audit logs, logging, monitoring, log export, workspace logs, subscriptions &
10 \\

Organizations &
organizations, orgs, enterprise, environments, site administration &
10 \\
\bottomrule
\end{tabularx}

\caption{Semantic categories derived from consensus clustering of OpenAPI tags. Representative tags are shown for each category. Only categories present in at least 10 APIs are retained.}
\label{tab:rq2_semantic_categories}
\end{table*}

\paragraph{(b) Semantic Category Analysis}
We now analyze coverage patterns across semantic categories derived from OpenAPI tags using the consensus BERT clustering procedure described in Stage~2. Table~\ref{tab:rq2_semantic_categories} reports the resulting semantic categories together with representative tags and their prevalence across the corpus. Table~\ref{tab:rq2_semantic_detailed} reports macro/micro coverage and lift for these categories.

\begin{table*}[t] 
\centering 
\small
\renewcommand{\arraystretch}{1.2}
\resizebox{\textwidth}{!}{
\begin{tabular}{lrrrrrrr} 
\toprule
\textbf{Semantic Category} &
\textbf{Macro Cov} &
\textbf{Macro Omit} &
\textbf{Macro Lift} &
\textbf{\# APIs} &
\textbf{Micro Cov} &
\textbf{Micro Omit} &
\textbf{\# Ops} \\
\midrule
Authorization 
& 0.0453 & \cellcolor{codebg!90!orange}\textbf{0.9547} & 0.3447 & 15 & 0.0574 & 0.9426 & 209 \\

Authentication 
& 0.0563 & \cellcolor{codebg!90!orange}\textbf{0.9437} & 0.2660 & 17 & 0.0547 & \cellcolor{codebg!90!orange}\textbf{0.9453} & 201 \\

Settings / configuration 
& 0.0960 & \cellcolor{codebg!90!orange}\textbf{0.9040} & 0.1536 & 11 & 0.0256 & \cellcolor{codebg!90!orange}\textbf{0.9744} & 117 \\

User management 
& 0.1125 & 0.8875 & 0.4011 & 17 & 0.0413 & \cellcolor{codebg!90!orange}\textbf{0.9587} & 242 \\

Key management 
& 0.1313 & 0.8687 & 0.5339 & 11 & 0.1250 & 0.8750 & 88 \\

Organizations 
& 0.1576 & 0.8424 & 1.1187 & 10 & 0.0691 & 0.9309 & 269 \\

Logs 
& 0.2100 & 0.7900 & \cellcolor{codeem!35}1.5018 & 10 & 0.0930 & 0.9070 & 43 \\

License / policy / requirements 
& 0.2267 & 0.7733 & 0.6093 & 16 & 0.1383 & 0.8617 & 188 \\

Data / storage 
& 0.2653 & 0.7347 & 1.3352 & 17 & 0.2529 & 0.7471 & 348 \\

Inventory management 
& 0.2914 & 0.7086 & 1.2702 & 20 & 0.2990 & 0.7010 & 311 \\

Event \& workflow management 
& 0.3362 & 0.6638 & \cellcolor{codeem!35}2.2388 & 17 & 0.0935 & 0.9065 & 460 \\

Analytics 
& 0.4107 & 0.5893 & \cellcolor{codeem!35}1.4960 & 15 & 0.7258 & 0.2742 & 186 \\

\midrule
\textbf{Median}
& \textbf{0.1838} & \textbf{0.8162} & \textbf{0.8640} & \textbf{15.5}
& \textbf{0.0933} & \textbf{0.9068} & \textbf{205.0} \\

\textbf{Q75}
& \textbf{0.2718} & \textbf{0.8916} & \textbf{1.3754} & \textbf{17.0}
& \textbf{0.1670} & \textbf{0.9433} & \textbf{279.5} \\

\bottomrule 
\end{tabular}
}
\caption{Semantic category coverage, omission, and lift across APIs (macro and micro aggregates). Cells highlighted in \textcolor{codeem}{blue} indicate high values (above the 75th percentile, Q75), while cells highlighted in orange indicate the highest omission values.}
\label{tab:rq2_semantic_detailed}
\end{table*}

\textbf{Operational and analytics categories are preferentially exposed as MCP tools.}
Table~\ref{tab:rq2_semantic_categories} lists the twelve semantic categories produced by consensus clustering. Event and workflow management, exhibits the strongest relative preference in the corpus (lift 2.24), followed by analytics (1.50), logs (1.50), data/storage (1.34), and inventory management (1.27). These categories share a common characteristic: they map to observability, orchestration, and data retrieval functionality. Analytics further stands out with the highest micro coverage of any category (0.73), confirming that its over-representation is not an artefact of per-API averaging.

\textbf{Security-sensitive and administrative categories are rarely exposed as MCP tools..}
At the opposite end, settings/configuration (lift 0.15), authentication (0.27), authorization (0.34), and user management (0.40) exhibit macro omission rates between 89\% and 95\% (Table~\ref{tab:rq2_semantic_detailed}). These categories govern credential handling, access control, and administrative configuration, functionality whose exposure through an LLM-facing interface carries inherent risk. 


\begin{figure}[h]
  \centering
  \includegraphics[width=0.99\linewidth]{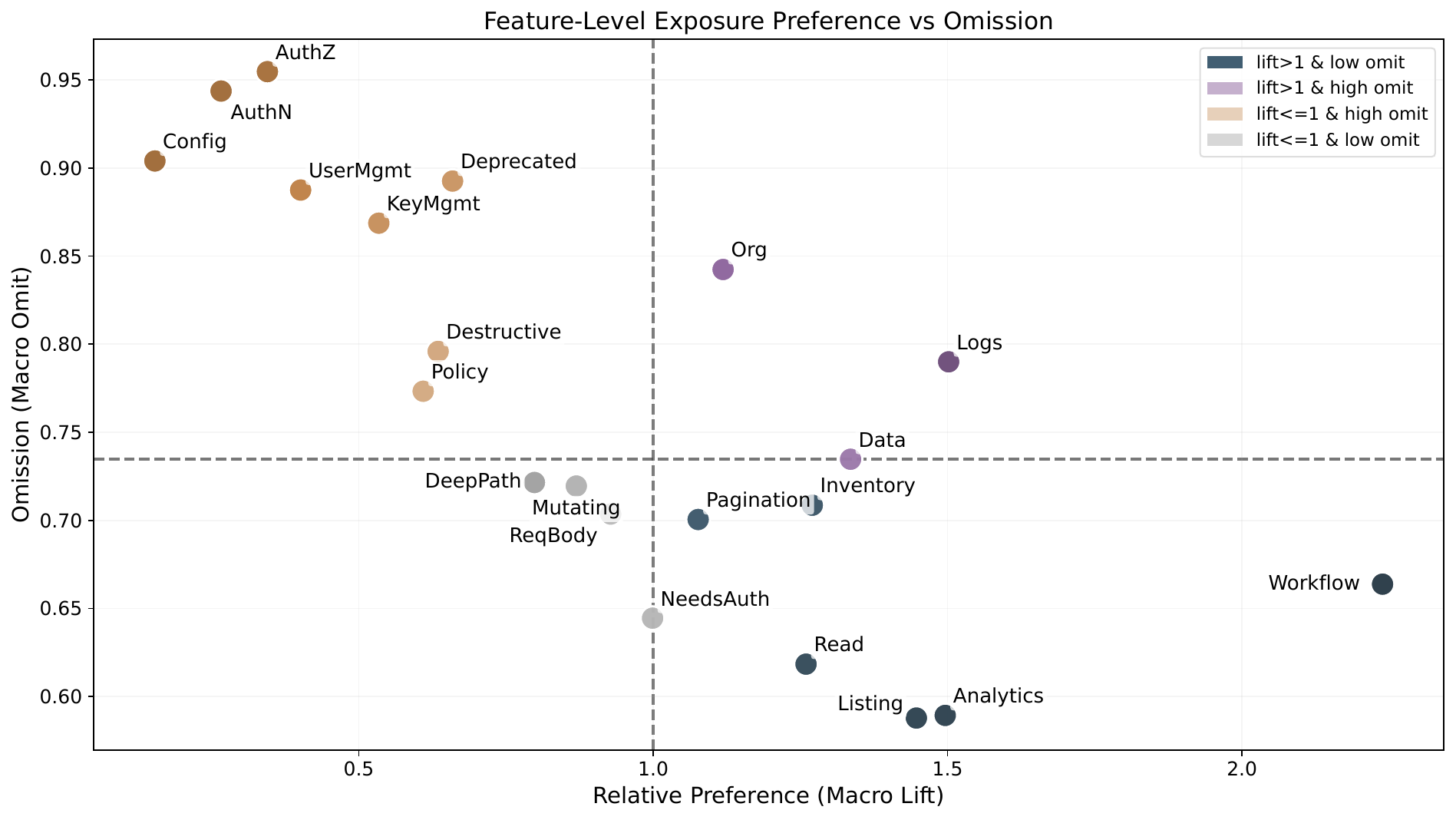}
\caption{Feature-level omission versus relative exposure preference.
Each point represents a feature category.
The vertical dashed line marks neutral preference (MacroLift = 1),
and the horizontal dashed line indicates the median macro omission (0.735),
separating lower- and higher-than-typical omission levels.}
  \label{fig:rq2_feature_quadrant}
\end{figure}

\textbf{Exposure patterns reflect active filtering rather than uniform downsampling.}
Figure~\ref{fig:rq2_feature_quadrant} plots macro omission against lift, combining both structural and semantic features. If MCP servers were downsampling operations uniformly, all points would cluster near lift $= 1$. Instead, the plot reveals a clear quadrant structure: retrieval and operational features concentrate in the bottom-right (preferred, lower omission), while security and administrative features occupy the top-left (disfavoured, higher omission). Structural complexity features, deep paths, request bodies, and mutating operations, fall near the centre, indicating proportional reduction rather than active filtering. This separation confirms that the coverage patterns reported above are the result of deliberate curation, not an artefact of API size.


\begin{findingbox}
\textbf{\faKey\ Finding-RQ2.3.} At the semantic level, observability, orchestration, and analytics categories are preferentially exposed, while authentication, authorization, and configuration categories exhibit near-complete omission. Quadrant analysis confirms that these inclusion and exclusion patterns are not a by-product of exposing a fixed fraction of each API, but reflect targeted selection based on operation semantics.
\end{findingbox}

\subsubsection{{RQ2.3: How Often Do MCP Tools Aggregate Multiple REST Operations?}}\mbox{}\
Following Stage~3, we characterize mapping structure from both the tool and operation perspectives. From the tool perspective, a tool is classified as \emph{one-to-one} if it aligns to exactly one REST operation and \emph{many-to-one} if it aligns to multiple operations. From the operation perspective, an operation is classified as \emph{one-to-many} if it is referenced by multiple tools. Table~\ref{tab:aggregation_patterns} summarises the observed mapping types, their prevalence, and representative examples from the corpus.

\begin{table*}[h]
\centering
\small
\resizebox{\linewidth}{!}{
\begin{tabular}{p{3.5cm} p{1.7cm} p{5cm} p{6cm}}
\toprule
\textbf{Mapping Type / \newline Pattern} &
\textbf{Tools -- APIs} &
\textbf{Description} &
\textbf{Example (Tool $\rightarrow$ Operation(s))} \\
\midrule
\rowcolor{gray!8}
\textbf{One-to-One} &
869/968 tools -- \newline
42 APIs &
A single MCP tool delegates to exactly one REST operation, preserving a direct
correspondence between the tool interface and the API endpoint. &
\texttt{get-source-languages} $\rightarrow$ \newline \texttt{GET /v2/languages} (DeepL)\cite{deepl_mcp} \\
\midrule
\multicolumn{4}{l}{\textbf{Many-to-One} \textit{(99/968 tools; 10.2\% across 32 APIs)}} \\
\midrule
Collection/Item Merge &
33/99\newline tools -- \newline
18 APIs &
Collapses the standard REST pair, a collection endpoint (\texttt{/resources})
and its corresponding item endpoint (\texttt{/resources/\{id\}}), into a single
tool with an optional identifier parameter. &
\texttt{workflows-get-default} $\rightarrow$ \newline \texttt{GET /api/v3/workflows}, \newline \texttt{GET /api/v3/workflows/\{id\}} \newline (Shortcut)\cite{shortcut_mcp} \\
\midrule
Multi-View Read &
20/99 \newline tools -- \newline
13 APIs &
Unifies three or more read-only endpoints that offer different views of the same
resource (e.g., list, detail, search, filter) into a single retrieval tool. &
\texttt{list\_users\_by\_org} $\rightarrow$ \newline \texttt{GET /orgs/\{id\}/users}, \newline \texttt{GET /org/users}, \newline \texttt{GET /orgs/\{id\}/users/search}, \newline \texttt{GET /org/users/lookup} \newline (Grafana)\cite{grafana_mcp} \\
\midrule
Lifecycle Action Bundle &
16/99 \newline tools -- \newline
9 APIs &
Groups operations that represent adjacent state transitions of the same resource
(e.g., assign, resolve, reopen), exposed through a single tool parameterised
by action type. &
\texttt{manage\_incident} $\rightarrow$ \newline \texttt{POST .../assign}, \newline \texttt{POST .../unassign}, \newline \texttt{POST .../resolve}, \newline \texttt{POST .../ignore}, \newline \texttt{POST .../reopen} \newline (GitGuardian)\cite{gitguardian_mcp} \\
\midrule
Scope Unification &
12/99 \newline tools -- \newline
8 APIs &
Merges global and tenant-scoped variants of the same operation into a single tool,
with a scope parameter selecting the target context. &
\texttt{get-v3-stats-total} $\rightarrow$ \newline \texttt{GET /v3/stats/total}, \newline \texttt{GET /v3/\{domain\}/stats/total}, \newline \texttt{GET /v3/stats/total/domains} \newline (Mailgun)\cite{mailgun_mcp} \\
\midrule
CRUD Collapse &
6/99 \newline tools -- \newline
4 APIs &
Combines create, read, update, and/or delete operations for a lightweight resource
into a single tool, parameterised by HTTP method or action. Observed only for
resources with simple, tightly coupled lifecycles. &
\texttt{manage\_repository\_notif\_subscription} $\rightarrow$ \newline \texttt{GET/PUT/DELETE /repos/\{o\}/\{r\}/subscription} \newline (Github)\cite{github_mcp} \\
\bottomrule
\end{tabular}}
\caption{Tool-to-operation mapping types and aggregation patterns observed across 42 matched APIs.}
\label{tab:aggregation_patterns}
\end{table*}

\textbf{Most MCP tools map to a single REST operation.}
Among aligned tools, 869 (89.8\%) map to exactly one REST operation, while only 99 (10.2\%) aggregate multiple operations (Table~\ref{tab:aggregation_patterns}). Aggregation appears in 32 of the 42 matched APIs; the remaining 10 rely exclusively on one-to-one mappings. This confirms that MCP tools predominantly act as direct projections of individual REST endpoints rather than composite abstractions.

\textbf{When multiple REST operations are aggregated into a single MCP tool (Many-to-one), the grouping follows a small number of recurring, well-motivated patterns.}
As Table~\ref{tab:aggregation_patterns} details, the five patterns differ in both frequency and abstraction strategy. The two most common, \emph{Collection/Item Merge} and\emph{Multi-View Read}, account for over half of all aggregated tools and share a retrieval-oriented motivation: both reduce the number of tools an LLM must select among when exploring a resource. The remaining three patterns address state management  \emph{Lifecycle Action Bundle}, scope resolution  \emph{Scope Unification}, and full lifecycle control  \emph{CRUD Collapse}; each appears in at least two distinct APIs, indicating shared design strategies rather than isolated decisions.

\textbf{Operation-level reuse (One-to-Many) is infrequent and concentrated on retrieval endpoints.}
Only 136 REST operations are referenced by more than one MCP tool. Reuse is modest (mean 2.37 tools per operation, median 2) and strongly skewed toward retrieval: 68\% of reused operations are \texttt{GET} endpoints, and roughly one-third involve listing operations. In 60.3\% of cases, the reusing tools share a common verb prefix (e.g., \texttt{get\_*}, \texttt{list\_*}), which is consistent with tools offering narrower, task-specific views of the same operation rather than introducing new functionality.
 
\begin{findingbox}
\textbf{\faKey\ Finding-RQ2.3.} MCP tools map to REST operations predominantly through one-to-one correspondences (89.8\%). When aggregation occurs, it follows five recurring patterns dominated by retrieval-oriented groupings. From the operation perspective, reuse is infrequent and similarly concentrated on read endpoints.
\end{findingbox}

\subsection{RQ3: Feasibility and Limitations of Automatic MCP Tool Generation from OpenAPI Specifications}

\subsubsection{Baseline Generation and Execution Success}

To assess whether OpenAPI specifications contain sufficient information for automatic MCP tool generation, we apply the AutoMCP baseline generator to all 77 REST-backed APIs in the corpus and evaluate the executability of the resulting tools. Table~\ref{tab:rq3-extended} reports per-API outcomes.

\textbf{Automatic generation from OpenAPI specifications succeeds for the majority of APIs, but failure case tend to affect entire APIs rather than individual tools.} At the API level, 59 of 77 APIs (76.6\%) achieve fully successful execution, every sampled tool invocation completes without error. The remaining 18 APIs (23.4\%) exhibit at least one failure. At the tool level, 1{,}666 of 2{,}190 sampled invocations succeed, yielding an overall execution success rate of 76\% (Table~\ref{tab:rq3-extended}). Notably, failures are not evenly spread: in 15 of the 18 failing APIs, failures affect all or most sampled tools, indicating that the root causes are API-wide properties, such as missing authentication schemes or incomplete server metadata, rather than individual endpoint defects.

\vspace{-6pt}
\begin{findingbox}
\textbf{\faKey\ Finding-RQ3.1.}
OpenAPI specifications are frequently sufficient for automatic MCP tool generation without manual integration code. When generation fails, the root causes are API-wide contract deficiencies rather than individual endpoint issues, suggesting that a small number of targeted specification improvements could substantially increase automation coverage.
\end{findingbox}
\vspace{-6pt}

\setlength{\tabcolsep}{2pt}
\renewcommand{\arraystretch}{1.05}

\begin{longtable}{@{}l c c c c c@{}}

\caption[Automation outcomes]{Automation outcomes for all 77 APIs (sorted by operation count).\newline
\footnotesize \textbf{Notes.}
Failure codes:
A: Incorrect or missing security schemes;
B: Malformed or relative base URLs;
C: Undocumented runtime headers and token prefixes;
D: Parameter type mismatches;
E: Missing endpoint authentication.}
\label{tab:rq3-extended} \\

\toprule
\textbf{API} & \textbf{Auth} & \textbf{Operations} & \textbf{Success} & \textbf{Failure} & \textbf{After Fix} \\
\midrule
\endfirsthead

\toprule
\textbf{API} & \textbf{Auth} & \textbf{Operations} & \textbf{Success} & \textbf{Failure} & \textbf{After Fix} \\
\midrule
\endhead

\midrule
\multicolumn{6}{r}{\footnotesize Continued on next page.} \\
\endfoot

\bottomrule
\endlastfoot

Brave & apikey & 1 & 1/1 & -- & -- \\
Duckduckgo & none & 1 & 1/1 & -- & -- \\
Weather & none & 1 & 1/1 & -- & -- \\
Advice & none & 3 & 2/2 & -- & -- \\
Globalping & OAuth2+http bearer & 4 & 4/4 & -- & -- \\
WebScraping.AI & apikey & 4 & 4/4 & -- & -- \\
Airtable & http bearer & 5 & 5/5 & -- & -- \\
Firecrawl & http bearer & 5 & 5/5 & -- & -- \\
IPLocate & apikey & 5 & 5/5 & -- & -- \\
Jobsoid & none & 6 & 6/6 & -- & -- \\
Adp & none & 7 & 0/7 & B & 7/7 \\
Exa & apikey & 7 & 7/7 & -- & -- \\
Exchange\_rate & none & 7 & 7/7 & -- & -- \\
Api Guru & none & 7 & 7/7 & -- & -- \\
Dropbox & OAuth2 & 9 & 9/9 & -- & -- \\
Red Hat Insights & apikey & 9 & 0/9 & A & 0/9 \\
Milvus & http bearer & 10 & 0/10 & B & 0/10 \\
Huggingface & apikey & 12 & 12/12 & -- & -- \\
VictoriaMetrics & apikey & 16 & 0/16 & B & 0/16 \\
Notion & http bearer & 19 & 1/19 & E+C & 1/19 \\
Petstore & none+apikey & 19 & 19/19 & -- & -- \\
Graphhopper & apikey & 20 & 20/20 & -- & -- \\

\midrule
2c2p & none & 22 & 2/2 & -- & -- \\
Ably & http basic+none & 22 & 11/11 & -- & -- \\
Token Metrics & none & 22 & 2/2 & -- & -- \\
Signwell & apikey & 23 & 5/5 & -- & -- \\
Keboola & apikey+none & 23 & 10/10 & -- & -- \\
Open\_route & apikey (query) & 24 & 2/8 & E & 2/8 \\
Pappers & apikey & 24 & 14/14 & -- & -- \\
Buttondown & http bearer & 28 & 0/21 & A+C & 21/21 \\
Todoist & http bearer & 28 & 15/15 & -- & -- \\
Apaleo & OAuth2 & 33 & 0/33 & A+B & 33/33 \\
Box & OAuth2 & 34 & 26/26 & -- & -- \\
Twiliomessaging & http basic & 40 & 6/6 & -- & -- \\
Qdrant & apikey & 43 & 0/17 & A & 17/17 \\
Innoship & apikey & 51 & 23/23 & -- & -- \\
Chatkitty & OAuth2 & 68 & 28/28 & -- & -- \\
Resend & http bearer & 70 & 17/17 & -- & -- \\
Gitlab & apikey & 73 & 19/23 & D & 19/23 \\
Meilisearch & http bearer & 73 & 23/23 & -- & -- \\
Modrinth & none+apikey & 76 & 44/44 & -- & -- \\
Multiversx & none & 76 & 17/17 & -- & -- \\
Circleci & apikey+http & 78 & 31/31 & -- & -- \\
Honeycomb & apikey+http bearer & 78 & 67/67 & -- & -- \\
Paddle & http bearer & 79 & 44/44 & -- & -- \\
Coingecko & apikey & 81 & 15/15 & -- & -- \\
Langfuse & http basic & 86 & 0/7 & B & 7/7 \\
Spotify & OAuth2 & 88 & 25/25 & -- & -- \\
Pokeapi & apikey+http & 97 & 48/48 & -- & -- \\
Elasticsearch & apikey & 102 & 8/8 & -- & -- \\
Neon & apikey+http bearer & 103 & 18/18 & -- & -- \\
Webflow & OAuth2+http bearer & 108 & 27/27 & -- & -- \\
Bitrise & apikey & 109 & 19/19 & -- & -- \\
Apify & http bearer & 120 & 44/44 & -- & -- \\
SonarQube & http bearer & 125 & 0/45 & A & 45/45 \\
ThoughtSpot & http bearer & 130 & 0/3 & C & 3/3 \\
Redis & apikey & 132 & 15/15 & -- & -- \\
Shortcut & apikey & 133 & 72/72 & -- & -- \\
Supabase & http bearer & 142 & 0/17 & B & 17/17 \\
Redis Cloud API & apikey & 142 & 34/34 & -- & -- \\
GitGuardian & http bearer & 159 & 60/60 & -- & -- \\
Render & http bearer & 166 & 69/69 & -- & -- \\
Mailgun & http basic & 176 & 58/58 & -- & -- \\
Sentry & http bearer & 183 & 18/18 & -- & -- \\
Twelve Data & apikey & 184 & 5/161 & -- & -- \\
Discord & OAuth2+apikey & 222 & 44/44 & -- & -- \\
Openai & http bearer & 226 & 43/43 & -- & -- \\
Trello & apikey+none & 255 & 55/55 & -- & -- \\
Mailchimp & http basic & 272 & 59/59 & -- & -- \\
Bitbucket & OAuth2+apikey+http & 318 & 31/31 & -- & -- \\
Square & OAuth2+none & 327 & 11/11 & -- & -- \\
PagerDuty & apikey & 386 & 108/108 & -- & -- \\
Confluent & OAuth2+http basic & 405 & 107/107 & -- & -- \\
MongoDB & apikey+OAuth2 & 462 & 0/6 & C & 6/6 \\
Clarifai & apikey & 489 & 0/73 & A+B+C & 0/73 \\
Stripe & http bearer & 583 & 0/151 & A & 151/151 \\
Github & OAuth2/http & 1114 & 0/81 & A & 81/81 \\

\midrule
\textbf{Total} & -- & \textbf{8{,}890} & \textbf{1{,}666/2{,}190} & -- & \textbf{2{,}063/2{,}190} \\

\end{longtable}

\subsubsection{Failure analysis and defect taxonomy}
\label{sec:rq3_failure_analysis}
While the baseline generator succeeds for the majority of APIs, Table~\ref{tab:rq3-extended} reveals that certain OpenAPI specifications consistently produce tools that fail at execution time. To understand what specification-level factors prevent faithful generation, we analyse all failed invocations using open coding (Section~\ref{subsec:rq3-approach}).

\textbf{Generation failures stem from discrepancies between the OpenAPI contract and the runtime behaviour of the API.}
Across the 18 APIs (23.4\%) that exhibit at least one execution failure, we identify five root causes through open coding (Table~\ref{tab:rq3-extended} and Table~\ref{tab:rq4-failure-summary}). The most prevalent is incorrect or missing security scheme declarations~(A), where the specification either omits the required authentication mechanism or declares one that differs from what the server enforces. Malformed or relative base URLs~(B) and undocumented runtime headers or token prefixes~(C) follow, each causing the generated tool to construct requests that the server cannot route or authenticate. Parameter type mismatches~(D) and missing endpoint-level authentication annotations~(E) are less frequent but produce the same effect: the generated request diverges from what the API expects at execution time. All five causes share a common characteristic, they arise not from limitations of the generator, but from incomplete or inaccurate specification of runtime requirements in the OpenAPI contract.

\begin{table}[H]
\centering
\caption{Root causes of automation failures observed in RQ3.}
\label{tab:rq4-failure-summary}
\small
\resizebox{0.82\textwidth}{!}{%
\begin{tabular}{@{}lcc@{}}
\toprule
\textbf{Failure Category} & \textbf{Affected APIs} & \textbf{Affected Tools (in sample)} \\
\midrule

\textbf{(A) Incorrect or missing security schemes}
& 8  & 18.9\% (416/2{,}190) \\

\textbf{(B) Malformed or relative base URLs}
& 7  & 6.8\% (149/2{,}190) \\

\textbf{(C) Undocumented runtime headers and token prefixes}
& 5  & 5.5\% (121/2{,}190) \\

\textbf{(D) Parameter type mismatches}
& 1  & 0.2\% (4/2{,}190) \\

\textbf{(E) Missing endpoint-level authentication annotations}
& 2  & 1.1\% (24/2{,}190) \\

\midrule
\textbf{Total}
& 18 & 23.9\% (524/2{,}190) \\

\bottomrule
\end{tabular}
}
\end{table}

Below, we discuss each category and illustrate the minimal contract change (when applicable)
that would be sufficient to eliminate the failure under the baseline generator.

\paragraph{\textbf{(A) Incorrect or missing security schemes.}}
In 8 APIs, authentication metadata is missing or incorrectly declared in the OpenAPI contract. Typical issues include absent entries under \texttt{components.securitySchemes}, deprecated OAuth2 flow declarations, or missing token endpoints. When such metadata is incomplete, AutoMCP cannot infer credential injection behavior, leading to systematic authorization failures at runtime. A minimal correction involves restoring the missing authentication fields required to complete the OAuth2 flow, as illustrated in Listing~\ref{lst:apaleo-sec}.

\begin{listing}[h]
\caption{Apaleo security scheme and patch (lines 7).}
\label{lst:apaleo-sec}
\begin{lstlisting}[style=code, language=diff]
 securitySchemes:
   oauth2:
     type: oauth2
     flows:
       authorizationCode:
         authorizationUrl: https://identity.[..]/authorize
+        tokenUrl: https://identity[..]/connect/token
\end{lstlisting}
\end{listing}

\paragraph{\textbf{(B) Malformed or relative base URLs.}}
Some specifications declare server URLs using unresolved templates or relative paths. Such declarations prevent correct request routing and result in immediate connectivity failures for all endpoints. In these cases, a minimal repair consists of replacing the placeholder with a fully qualified base URL, as shown in Listing~\ref{lst:adp-base}.

\begin{listing}[h]
\caption{ADP base URL placeholder replaced with an absolute URL. (line 2).}
\label{lst:adp-base}
\begin{minipage}{0.85\linewidth}
\begin{lstlisting}[style=code, language=diff]
Servers:
-   url: '{{service-root}}'
+   url: 'https://api.adp.com'
\end{lstlisting}
\end{minipage}
\end{listing}

\paragraph{\textbf{(C) Undocumented runtime headers and token prefixes.}}
Some APIs require metadata not captured in the OpenAPI specification, such as version-specific headers or non-standard token formats (e.g., a mandatory \texttt{Bearer} prefix). Because this information appears only in prose documentation, the generator cannot infer it, producing requests that are structurally valid but rejected by the server. Declaring these requirements as explicit header parameters or security scheme properties in the specification would close the gap.

\begin{listing}[h]
\centering
\caption{Notion: adding required \texttt{Notion-Version} header to align specification with runtime requirements.}
\label{lst:notion-header}
\begin{minipage}{0.75\linewidth}
\begin{lstlisting}[style=code, language=diff]
/v1/users/me:
  get:
    summary: Retrieve your token's bot user
    [...]
   parameters:
       [...]
+     - name: Notion-Version
+       in: header
+       required: true
+       schema:
+         type: string
+         default: "2022-06-28"
+       description: The Notion API version to use
\end{lstlisting}
\end{minipage}
\end{listing}

\paragraph{\textbf{(D) Parameter type mismatches.}}
 A small fraction of failures stem from mismatches between parameter types declared in the OpenAPI schema and those accepted by the live API. For instance, identifiers declared as integers may be treated as opaque strings in practice. Correcting the parameter schema to reflect the actual accepted type is sufficient to restore correct request construction, as illustrated in Listing~\ref{lst:gitlab-param}.

\begin{listing}[h]
\centering
\caption{GitLab: changing project id from \texttt{integer} to \texttt{string}.}
\label{lst:gitlab-param}
\begin{minipage}{0.65\linewidth}
\begin{lstlisting}[style=code, language=diff]
parameters:
    name: id
    [...]
    schema:
-       type: integer
+       type: string
\end{lstlisting}
\end{minipage}
\end{listing}

\paragraph{\textbf{(E) Missing endpoint-level authentication.}} Some specifications explicitly mark certain operations as public (\texttt{security: []}) even though the server enforces authentication on them at runtime. The generator trusts the contract and omits credentials from these requests, which the server then rejects. Removing the incorrect public override and allowing the global security scheme to apply would resolve these cases. A potential repair involves inheriting the global security scheme for any operation that does not declare its own, as illustrated in Listing~\ref{lst:notion-security}.

\begin{listing}[h]
\centering
\caption{Notion: removing incorrect endpoint-level security override to inherit global Bearer authentication.}
\label{lst:notion-security}
\begin{minipage}{0.75\linewidth}
\begin{lstlisting}[style=code, language=diff]
# Global security (applies to all endpoints)
security:
  - bearerAuth: []
 [...]
/v1/pages/{page_id}:
  get:
    summary: Retrieve a page
    [...]
-   security: []
\end{lstlisting}
\end{minipage}
\end{listing}

\vspace{-6pt}
\begin{findingbox}
\textbf{\faKey\ Finding-RQ3.2.}
Fully automated generation of MCP servers from OpenAPI specifications fails due to \textbf{recurring, specification-level
defects} that systematically impact many tools within an API. These deficiencies are identifiable through static analysis of the openAPI contract and correctable through targeted edits, suggesting that specification quality, is the primary bottleneck for automation scalability.
\end{findingbox}
\vspace{-6pt}

\subsection{RQ4: Effectiveness of Automated Repairs and Tool-Set Transformations (filtering and grouping) for OpenAPI-Driven MCP Generation}
\subsubsection{Specification Defect Detection and Repair}

We evaluate SpecFix (Section~\ref{subsec:rq4-approach}) in three stages: detection accuracy on validated defect instances, correctness of the generated patches, and end-to-end impact on generation outcomes across the full corpus.

\textbf{SpecFix detects targeted specification defects with high recall and precision.}
Of the 18 APIs with confirmed failures in RQ3 (Table~\ref{tab:rq3-extended}), 16 contain at least one defect in the categories targeted by SpecFix: authentication misconfigurations~(A), malformed base URLs~(B), and undocumented runtime headers or token prefixes~(C). These 16 APIs comprise 20 defect instances in total. The remaining 2 APIs exhibit only out-of-scope defects (one~D, one~E) and are excluded from the detection and repair evaluation. Table~\ref{tab:rq4_specfix} reports outcomes for both stages. SpecFix correctly identifies 19 of the 20 targeted instances (recall 95\%), with perfect detection for categories A~(8/8) and B~(7/7), and one missed instance in category~C~(4/5) where the required header appeared only in external prose inaccessible to the analyser. The tool produced a single false positive, an A-type flag for a specification declaring both bearer and basic authentication where the documentation referenced only bearer, yielding an overall precision of 95\%.

\textbf{SpecFix produces correct repairs for the majority of detected defects, with failures confined to cases with insufficient documentation evidence.}
All 19 patches produced from true detections result in parseable OpenAPI specifications, confirming that the repair pipeline preserves structural validity. However, three patches are incorrect (Table~\ref{tab:rq4_specfix}): two involve base-URL defects~(B) where the LLM defaulted to a placeholder (\texttt{https://api.example.com}) because no recoverable endpoint was present in the documentation, and one involves an authentication defect~(A) where the API expects a custom identity header but the repair replaced it with a standard HTTP bearer scheme. The overall repair correctness rate is 16/19 (84.2\%). These cases illustrate a current limitation: when documentation evidence is insufficient or ambiguous, the LLM generates plausible but incorrect patches rather than abstaining.

\textbf{Most APIs with targeted defects achieve full recovery after repair.}
Of the 16 APIs evaluated, 11 (68.8\%) achieve full recovery, every previously failing invocation now succeeds (Table~\ref{tab:rq4_specfix}, Table~\ref{tab:rq3-extended}). Two APIs achieve partial recovery: one contains an additional out-of-scope defect~(E) that persists after the targeted C-type defect is repaired, and the other has its A and B defects corrected but a co-occurring C-type defect that was not detected. The remaining three APIs are not fixed due to the three incorrect repairs identified above (one~A, two~B).

\textbf{At the corpus level, specification repair substantially improves end-to-end generation success.}
Table~\ref{tab:rq4_endtoend} reports end-to-end outcomes after rerunning the baseline pipeline (Section~\ref{subsec:rq4-approach}) on all 77 APIs with SpecFix patches applied. At the API level, fully successful generation increases from 59/77 (76.6\%) to 67/77 (87.0\%). At the tool level, successful invocations rise from 1{,}666/2{,}190 (76.0\%) to 2{,}063/2{,}190 (94.2\%), an improvement of 18.2 percentage points. The disproportionate tool-level gain reflects the concentrated nature of failures: repairing a single API-wide defect restores execution for all of its sampled tools simultaneously.

\textbf{Repair introduces a small number of regressions on previously passing APIs.}
SpecFix produced regressions in 3 of the 59 previously passing APIs, affecting 10 tool invocations (0.45\% of all sampled tools). In each case, SpecFix inferred an authentication requirement from documentation fragments for endpoints that were either publicly accessible or required credentials only under specific conditions. These over-constrained patches highlight a trade-off in documentation-driven repair: when the specification is ambiguous about authentication scope, the repair logic defaults to the more restrictive interpretation.


\begin{table}[t]
\centering
\small
\caption{SpecFix pipeline outcomes: detection and repair quality across targeted defect categories (20 defect instances across 16 APIs).}
\label{tab:rq4_specfix}
\begin{tabular}{l rr rr rr}
\toprule
& \multicolumn{4}{c}{\textbf{Detection}} & \multicolumn{2}{c}{\textbf{Repair}} \\
\cmidrule(lr){2-5} \cmidrule(lr){6-7}
\textbf{Defect category} &
\textbf{Detected} & \textbf{FP} & \textbf{Recall} & \textbf{Precision} &
\textbf{Correct} & \textbf{Rate} \\
\midrule
Auth.\ misconfiguration (A) &
8/8 & 1 & 100\% & 88.9\% &
7/8 & 87.5\% \\
Malformed base URL (B) &
7/7 & 0 & 100\% & 100\% &
5/7 & 71.4\% \\
Undoc.\ header/token prefix (C) &
4/5 & 0 & 80\% & 100\% &
4/4 & 100\% \\
\midrule
\textbf{Total (A--C)} &
\textbf{19/20} & \textbf{1} & \textbf{95\%} & \textbf{95\%} &
\textbf{16/19} & \textbf{84.2\%} \\
\midrule
\multicolumn{7}{l}{\textbf{Recovery outcomes (16 APIs with A--C defects)}} \\
\midrule
\multicolumn{5}{l}{Full recovery} &
\multicolumn{2}{r}{11/16 \quad (68.8\%)} \\
\multicolumn{5}{l}{Partial recovery (out-of-scope or undetected defect persists)} &
\multicolumn{2}{r}{2/16 \quad (12.5\%)} \\
\multicolumn{5}{l}{Not fixed (incorrect repair)} &
\multicolumn{2}{r}{3/16 \quad (18.8\%)} \\
\bottomrule
\end{tabular}
\end{table}

\begin{table}[t]
\centering
\small
\caption{End-to-end generation outcomes before and after SpecFix (77 APIs).}
\label{tab:rq4_endtoend}
\begin{tabular}{l rr r}
\toprule
\textbf{Metric} & \textbf{Baseline} & \textbf{After repair} & \textbf{Improvement} \\
\midrule
APIs with full execution success & 59/77 (76.6\%) & 67/77 (87.0\%) & +8 (+10.4 pp) \\
Successful tool invocations & 1{,}666/2{,}190 (76.0\%) & 2{,}063/2{,}190 (94.2\%) & +397 (+18.2 pp) \\
Regressions (previously passing) & -- & 3/59 APIs (5.1\%) & 10 tools affected \\
\bottomrule
\end{tabular}
\end{table}

\begin{findingbox}
\textbf{\faKey\ Finding-RQ4.1.}
Automated specification repair raises tool-level execution success from 76\% to 94\%, with the strongest gains on localised, well-documented defects. Residual failures arise from defect classes outside the current repair scope and from ambiguities in authentication semantics, where documentation-driven inference can produce over-constrained patches.
\end{findingbox}

\subsubsection{Tool Filtering and Operation Grouping}

Repairing specification defects improves execution reliability but does not address a second practical barrier: tool-space explosion. Large APIs can produce hundreds of generated tools (Table~\ref{tab:rq3-extended}), exceeding the context-window and tool-selection capacity of current LLMs. We therefore apply two complementary transformations, category-based filtering and deterministic grouping, grounded in the coverage and aggregation patterns identified in RQ2 (Section~\ref{subsec:rq4-approach}).

\textbf{Category-based filtering removes nearly one fifth of the generated tool space.}
As illustrated in \autoref{tab:rq4_reduction_summary}, across the 77 APIs, filtering removes 1{,}636 of 8{,}890 candidate operations ($-$18.4\%). The per-API impact varies with API composition: the mean reduction is 21.2 operations (median 5).

\textbf{Grouping preserves behavioural equivalence.} All 232 endpoints evaluated under the Collection/Item Merge pattern passed both invocation modes, confirming that merged tools reproduce the behaviour of their original counterparts without regressions.

\textbf{Collection/Item Merge grouping removes 486 tools across 55 APIs.}
After filtering, we apply the Collection/Item Merge pattern (Table~\ref{tab:aggregation_patterns}) to consolidate list and detail endpoint pairs (e.g., \texttt{GET /items} and \texttt{GET /items/\{id\}}) into single tools with an optional identifier parameter. Across the corpus, 55 APIs contain mergeable patterns, yielding 485 merge groups that remove an additional 486 tools, reducing the post-filtering total from 7{,}254 to 6{,}768.

\begin{figure}[h]
\centering
\begin{subfigure}{0.48\linewidth}
    \centering
    \includegraphics[width=\linewidth]{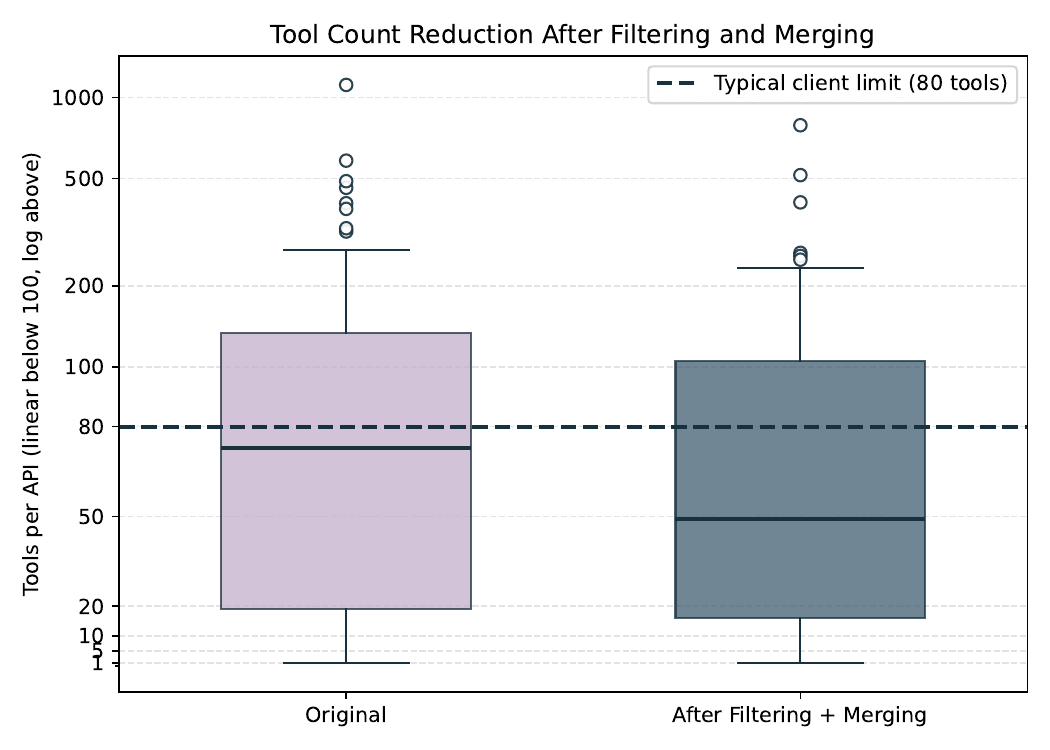}
    \caption{Distribution of tools per API before and after reduction.}
    \label{fig:rq4_boxplot}
\end{subfigure}
\hfill
\begin{subfigure}{0.48\linewidth}
    \centering
    \includegraphics[width=\linewidth]{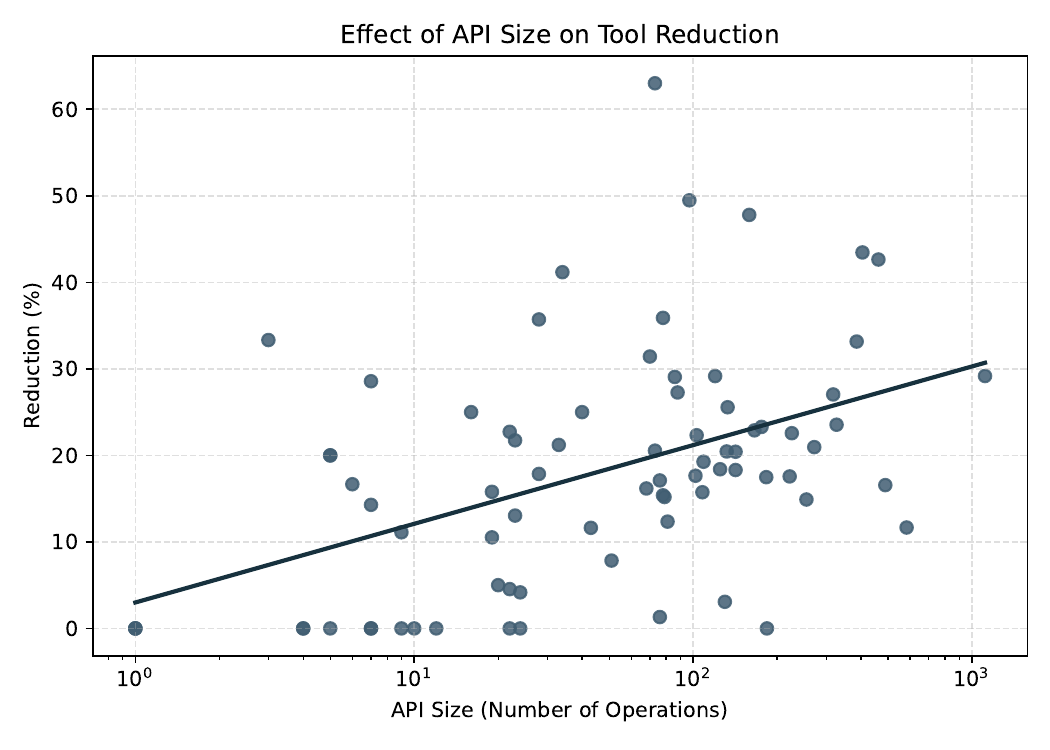}
    \caption{Per-API reduction as a function of original API size.}
    \label{fig:rq4_scatter}
\end{subfigure}
\caption{Effect of category-based filtering and Collection/Item Merge grouping on tool-space complexity.}
\label{fig:rq4_reduction}
\end{figure}

\textbf{Together, the two transformations reduce the median tool count by one third.}
Table~\ref{tab:rq4_reduction_summary} reports the cumulative effect. The total tool count decreases from 8{,}890 to 6{,}768 ($-$23.9\%), and the median per-API count drops from 73 to 49 ($-$32.9\%). Figure~\ref{fig:rq4_boxplot} shows that both the median and interquartile range shift downward, indicating consistent reductions across APIs rather than gains driven by a few outliers. While these reductions are substantial, they remain higher than the level of abstraction observed in real MCP servers, where developers often reduce hundreds of operations to a much smaller set of tools. This gap reflects a deliberate design choice: our transformations focus on operations that can be safely omitted or merged based on structural and semantic criteria, without introducing assumptions about user intent. More aggressive reductions likely require reasoning about user-facing functionality and LLM usability, such as identifying task-oriented abstractions or high-level workflows. We leave this direction to future work.

\textbf{Larger APIs benefit disproportionately from the reduction pipeline.}
Figure~\ref{fig:rq4_scatter} plots per-API reduction against original API size. APIs with larger initial tool sets exhibit higher absolute reductions. The largest absolute reductions occur in GitHub ($-$325 tools), MongoDB ($-$197), and Confluent ($-$176), while the highest relative reductions appear in Meilisearch (63.0\%), PokeAPI (49.5\%), and GitGuardian (47.8\%).


\begin{table}[t]
\centering
\small
\caption{Tool-space reductions after category-based filtering and Collection/Item Merge grouping (77 APIs).}
\label{tab:rq4_reduction_summary}
\begin{tabular}{lrr}
\toprule
\textbf{Metric} & \textbf{Value} & \textbf{Change} \\
\midrule
Total tools (baseline) & 8{,}890 & -- \\
After category-based filtering & 7{,}254 & $-$18.4\% \\
After filtering + grouping & 6{,}768 & $-$23.9\% \\
\midrule
Median tools per API (baseline) & 73 & -- \\
Median tools per API (after filtering) & 60 & $-$17.8\% \\
Median tools per API (after filtering + grouping) & 49 & $-$32.9\% \\
\bottomrule
\end{tabular}
\end{table}



\begin{findingbox}
\textbf{\faKey\ Finding-RQ4.2.}
Empirically grounded filtering and grouping, reduce the generated tool space by 23.9\% and lower the median per-API tool count from 73 to 49. Reductions are consistent across the distribution and most pronounced for large APIs, bringing tool sets closer to the practical limits of current LLM-based agents.
\end{findingbox}

%% file: Sections/Limitations.tex
\section{Threats to Validity}
\label{sec:threats-to-validity}

We discuss threats to validity following established guidelines for empirical software engineering research~\cite{yin2009case}.

\textbf{Internal validity.}
Several analysis stages rely on manual coding (REST reliance classification, tool--operation alignment, failure root-cause labelling, and repair validation). To control for subjectivity, all tasks were performed independently by at least two authors using shared protocols, with inter-rater reliability measured before reconciliation (Cohen's $\kappa$: 0.87--1.0; Krippendorff's $\alpha$: 0.81--1.0). Disagreements were resolved through code-grounded evidence review.

In RQ2, we use GPT-4.1 to scale tool--operation alignment beyond a 10-API gold standard. The model achieved macro-$F_1 = 0.87$ and exact-match accuracy of 0.78 on the gold set, and the prompt was applied without modification to remaining servers; nonetheless, residual labelling errors may affect coverage statistics. We report both macro and micro aggregates to reduce the influence of individual APIs.

Tool execution in RQ3--RQ4 is subject to transient upstream conditions (rate limiting, downtime, stale state). We mitigated this through dependency-aware ordering, re-execution of suspected transient failures, and independent re-evaluation of a random 10\% subset ($\kappa = 0.97$; $\alpha = 1.0$). Additionally, SpecFix intentionally targets only three of five defect categories; the reported repair effectiveness therefore represents a lower bound.

\textbf{External validity.}
Our RQ1 corpus comprises 116 servers drawn from Anthropic's official MCP catalogue (July 2025), filtered to repositories with a publicly documented REST API and $\geq$10 GitHub stars. This selection captures mature, officially listed servers but may not generalise to private, enterprise-internal, or community-developed deployments.

%% file: Sections/Conclusion.tex
\section{Conclusion}
\label{sec:conclusion}

This paper presented the first empirical study of how MCP servers relate to vendor REST APIs and whether this relationship can be exploited for automated server generation. Through a multi-stage analysis of 116 official MCP servers, 42 paired OpenAPI specifications, and an extended corpus of 77 real-world API contracts, we addressed four complementary research questions spanning architecture characterisation, operation-level mapping, generation feasibility, and automated repair.

Our findings paint a consistent picture. The vast majority of official MCP servers are REST-backed, and most implement tools as thin wrappers that perform bare API consumption rather than complex orchestration. Yet these servers expose only a selective fraction of the underlying API surface, typically around one-fifth of documented operations, following systematic omission and grouping patterns that are structurally and semantically predictable from the OpenAPI specification alone.

This characterisation motivated and informed an automated generation pipeline. We showed that a fixed, specification-driven translation already produces executable MCP tools for 76\% of sampled operations, and that the remaining failures concentrate in a small number of mechanically identifiable specification defect classes. SpecFix, our automated repair component, raised tool-level success to 94.2\% by targeting the three most prevalent defect categories, while empirically grounded filtering and regrouping reduced the median tool count per API by roughly one-third, mitigating the tool-set explosion that degrades LLM tool selection at scale.

For practitioners, the study provides two concrete resources that are not currently available in the MCP ecosystem. First, the omission and grouping patterns documented in RQ2 constitute empirically grounded design heuristics for MCP server construction, offering a reference for which operation categories to expose, which to suppress, and how to structure tool interfaces, based on what experienced developers consistently do across dozens of independently authored servers. Second, the defect taxonomy from RQ3 gives API providers a targeted checklist of specification issues that silently break downstream automation.